\documentclass[%
 aip,
 jmp,%
 amsmath,amssymb,
 reprint,%
]{revtex4-2}

\usepackage{graphicx}
\usepackage{subcaption}
\usepackage{parselines}
\usepackage[export]{adjustbox}
\usepackage{lipsum}
\usepackage{url}
\usepackage{dcolumn}
\usepackage{bm}
\usepackage{booktabs}
\usepackage{comment} 
\usepackage[utf8]{inputenc}
\usepackage[T1]{fontenc}
\usepackage{mathptmx}
\usepackage{etoolbox}
\makeatletter
\def\@email#1#2{%
 \endgroup
 \patchcmd{\titleblock@produce}
  {\frontmatter@RRAPformat}
  {\frontmatter@RRAPformat{\produce@RRAP{*#1\href{mailto:#2}{#2}}}\frontmatter@RRAPformat}
  {}{}
}%
\makeatother

\begin{document}

\preprint{AIP/123-QED}

\title{The Realization of a Gas Puff Imaging System on the Wendelstein 7-X Stellarator}

\author{J.L. Terry}%
    \email{terry@psfc.mit.edu}
\affiliation{ 
Massachusetts Institute of Technology - Plasma Science and Fusion Center, Cambridge, MA 02139, USA
}%

\author{A. von Stechow}
\affiliation{ 
Max-Planck-Institut für Plasmaphysik, Greifswald, Germany
}%

\author{S.G. Baek}
\affiliation{ 
Massachusetts Institute of Technology - Plasma Science and Fusion Center, Cambridge, MA 02139, USA
}%

\author{S.B. Ballinger}
\affiliation{ 
Massachusetts Institute of Technology - Plasma Science and Fusion Center, Cambridge, MA 02139, USA
}%

\author{O. Grulke}
\altaffiliation[Also at ]{Department of Physics, Technical University of Denmark, Lyngby, Denmark
}%

\author{C. von Sehren}
\author{R. Laube}
\author{C. Killer}
\author{F. Scharmer}

\author{K.J. Brunner}
\affiliation{ 
Max-Planck-Institut für Plasmaphysik, Greifswald, Germany
}%

\author{J. Knauer}
\affiliation{ 
Max-Planck-Institut für Plasmaphysik, Greifswald, Germany
}%

\author{S. Bois}
\affiliation{ 
Laboratoire de Physique des Plasmas, Ecole Polytechnique-CNRS-Univ Paris-Sud-UPMC, Palaiseau, France
}%

\author{the W7-X Team}
\affiliation{ 
Max-Planck-Institut für Plasmaphysik, Greifswald, Germany
}%

\date{\today}

\begin{abstract}
A system for studying the spatio-temporal dynamics of fluctuations in the boundary of the W7-X plasma using the ``Gas-Puff Imaging'' (GPI) technique has been designed, constructed, installed, and operated. This GPI system addresses a number of challenges specific to long-pulse superconducting devices like W7-X, including the long distance between the plasma and the vacuum vessel wall, the long distance between the plasma and diagnostic ports, the range of last closed flux surface (LCFS) locations for different magnetic configurations in W7-X, and management of heat loads on the system's plasma-facing components. The system features a pair of ``converging-diverging'' nozzles for partially collimating the gas puffed locally $\approx$135 mm radially outboard of the plasma boundary, a pop-up turning mirror for viewing the gas puff emission from the side (which also acts as a shutter for the re-entrant vacuum window), and a high-throughput optical system that collects visible emission resulting from the interaction between the puffed gas and the plasma and directs it along a water-cooled re-entrant tube directly onto the 8 x 16 pixel detector array of the fast camera. The DEGAS 2 neutrals code was used to simulate the H$_\alpha$ (656 nm) and the HeI (587 nm) line emission expected from well-characterized gas-puffs of H$_2$ and He and excited within typical edge plasma profiles in W7-X, thereby predicting line brightnesses used to reduce the risks associated with system sensitivity and placement of the field of view. Operation of GPI on W7-X shows excellent signal to noise ratios ($>$100) over the field of view for minimally perturbing gas puffs. The GPI system provides detailed measurements of the 2-dimensional (radial and poloidal) dynamics of plasma fluctuations in the W7-X edge and scrape-off layer, and in and around the magnetic islands outside the LCFS that make up the island divertor configuration employed on W7-X. 

\end{abstract}

\maketitle


\section{\label{sec:level1}Introduction}
The understanding of edge and scrape-off layer (SOL) transport in magnetically confined plasmas continues to be of crucial importance in the development of a viable fusion power plant. It affects critical aspects of energy and particle exhaust handling, e.g., divertor and first-wall lifetimes, and fuel recycling, as well as critical aspects of core plasma performance (via the pedestal) and fueling. It has long been recognized that turbulence is important to edge transport, and the past two decades have seen significant advances in our understanding of edge phenomena in tokamaks and stellarators. These advances have been made possible by the combination of advanced theoretical/computational models and detailed diagnostic information. In particular, a number of imaging diagnostic systems yielding 2-dimensional information have been developed for the visualization of turbulent dynamics, including Beam Emission Spectroscopy (BES) \cite{paul1990neutral,McKee1999}, Electron Cyclotron Emission Imaging (ECEI) \cite{cima1997ece,deng2001electron,tobias2010commissioning}, Microwave Imaging Reflectometry (MIR) \cite{mazzucato2001microwave,munsat2003microwave}, and (most importantly in the context of this report) Gas-Puff-Imaging (GPI) \cite{terry2001visible,maquedaGPI2001,zweben_GPI_NSTX_2004}. GPI is a mature diagnostic technique which has been of particular importance for identifying the structure and dynamics of edge phenomena in tokamaks: Alcator C-Mod \cite{cziegler2010experimental}, NSTX \cite{zweben_GPI_NSTX_2004}, ASDEX-Upgrade \cite{Fuchert_2014}, TEXTOR \cite{shesterikov_GPI_TEXTOR_2013}, EAST \cite{liu_EAST_GPI_2012}, and TCV \cite{offeddu2022gas}, and in a reversed-field-pinch -  RFX-Mod \cite{agostini_GPI_RFX_2006}  - see review of GPI by Zweben, et al. \cite{Zweben_invited2017}. This report describes the design, component characterization, and performance of a GPI system deployed on the Wendelstein 7-X (W7-X) stellarator \cite{klinger2019overview,wolf2019performance,grulke_iaea_overview_2023}. 

GPI has proven to be valuable for understanding a number of key physics issues of the plasma boundary and the SOL in tokamaks, including but not limited to the following examples: the dynamics of highly intermittent fluctuations in the SOL, often referred to as filaments or blobs \cite{zweben2016blob,Fuchert_2014}, the dominant role of these filaments in the far-SOL, see e.g. \cite{militello2016scrape}, the development of models for filament intermittency and filament dynamics \cite{garcia2013intermittent,DIppolito_comp_with_theory_2011}, dynamics during the L-H transition \cite{Diallo_2017,xu_GPI_L-H_study_2013}, SOL dynamics during ELMs \cite{GPI_during_ELMs_Lampert_2021}, the role that filamentary radial transport plays in the tokamak density limit \cite{terry2005transport}, the nonlinear turbulent kinetic energy transfer from the background turbulence into sheared quasi-static flows \cite{cziegler2014zonal}, zonal flows and geodesic acoustic modes \cite{cziegler2013fluctuating}, and the characterization of coherent electromagnetic modes \cite{cziegler2010experimental}.  

While filaments are typically present in the W7-X SOL, their transport and intermittency statistics are very different from what is found in tokamaks \cite{Zoletnik_W7X_filaments, Killer_filaments_PPCF_2020}. The presence of edge magnetic islands in W7-X appears to play a role in the filament dynamics, as does the field-line averaged curvature \cite{Zoletnik_W7X_filaments, Killer_filaments_PPCF_2020,huslage2023coherence}. 2D imaging by GPI at the outboard island in W7-X will help to elucidate those dynamics and the effects of the islands on the filaments \cite{killer2021turbulent}. Regarding edge coherent modes in W7-X, a number have been observed \cite{liu_QCM_W7X_2018, Han_QCM_W7-X_NF_2019,Zoletnik_W7X_filaments,kramer2020_SOL_investigation,ballinger2021_1kHz}, and GPI will further characterize and investigate those modes. 

\subsection{\label{sec:level1}Key components of Gas-Puff Imaging}

\begin{figure}
    \center
    \includegraphics[scale=0.3]
    {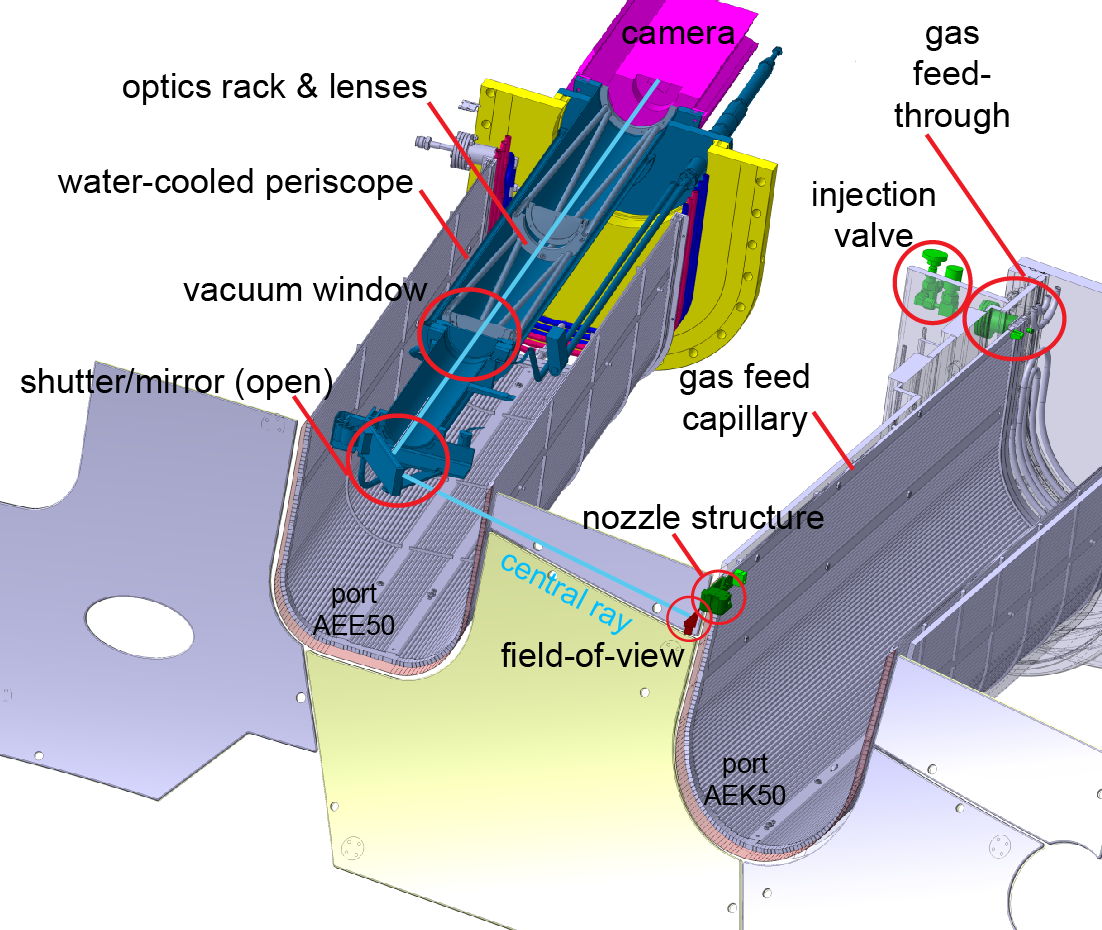}
    \caption{CAD drawing of the GPI system on W7-X. The plasma (not shown) occupies the bottom left of the drawing. The ``gas puff'' component is located in the horizontal vessel access port on the right. The ``light collection'' component is in the water-cooled re-entrant immersion tube in the port on the left. The ``fast camera'' component (in pink) is mounted at the end of the tube just beyond the large port flange (in yellow). The distance from this flange to the plasma vessel wall is $\approx$1.9 m, and from there to the LCFS of the ``standard'' W7-X magnetic field at the location of the nozzle is $\approx$0.17 m.}
   \label{fig:GPI_CAD}
 \end{figure}

The GPI technique is well-described in the GPI review publication \cite{Zweben_invited2017}. We reiterate the essentials and key components of the technique here. A “gas puff" component provides a source of hydrogenic or helium atoms with a limited toroidal extent at the edge of the plasma. (A vertical “sheet" of atoms is the ideal.) As the atoms from the puff penetrate the edge plasma, line emission is excited by the local plasma electrons. Under conditions \cite{AMathews_GPI_RSI} typically met in the edge plasmas of stellarators and tokamaks, the line emission from $H_\alpha$ (when puffing H$_2$) or the $3^3D - 2^3P$ transition of HeI at 587 nm (when puffing He) will represent excitation by a unique $n_e$ and $T_e$ in the local plasma and therefore also the combined effects of fluctuations in those quantities. The line emission is collected from the region “illuminated" by the gas puff and conveyed, by the “light collection" component, to a fast-framing 2-dimensional detector, the “fast camera" component. The viewing geometry is important. Ideally, the viewing chords should be aligned with the magnetic field lines at the gas puff location since the fluctuations are primarily field-aligned and the most interesting viewing field is therefore perpendicular to them. Any toroidal extent of the emission from the gas cloud will limit the spatial resolution in the images compared to that obtainable using the ideal ``sheet'' illumination. While the GPI concept is relatively simple, the realization of an actual system on a magnetic-confinement fusion-relevant device like W7-X is quite challenging as will be discussed below. The CAD model of these GPI system components, as realized on W7-X, is shown in Figure \ref{fig:GPI_CAD}.

\subsection{\label{sec:W7-X challenges}Challenges for GPI on W7-X}
W7-X is a large superconducting stellarator which provides plasma parameters and engineering demonstrations relevant to a future stellarator power plant. Its design was based on optimizing seven criteria, one of which is reduced neoclassical transport of the thermal plasma \cite{wolf2019performance}. It is developing the capability to maintain plasma durations of thirty minutes, and plasmas of 8 minutes duration and over 1 GJ of input (and exhaust) energy have already been realized \cite{grulke_iaea_overview_2023}. W7-X can accommodate a variety of magnetic configurations and utilizes the magnetic island divertor concept for heat and particle exhaust handling \cite{Gao_2019}. Elucidation of the islands' role in SOL transport is a primary goal for GPI investigations. These realities present significant challenges for implementing GPI on W7-X. As noted above, the ``gas puff'' component must be in-vessel and close to the plasma boundary. Ports with tangential views of the plasma that fulfill the requirement of field-aligned viewing chords are too small for the light collection requirements. Therefore, the “light collection" component must enter from one of the large horizontal ports and have parts relatively close to the plasma. These port flanges are roughly {2\,m} radially from the plasma edge in order to accommodate the coils and cryostat. Furthermore, the variety of magnetic configurations created in W7-X results in a range of last closed flux surface (LCFS) locations and island geometries that affect the decisions about the placements of the nozzle and field of view (FOV). We summarize the design challenges for GPI \textit{on W7-X} below. We believe that each represents a challenge not faced by any current or previously existing GPI system.
\begin{itemize}
    \item The variety of available SOL and island locations in the W7-X configuration space requires a large distance ($\sim$0.1 m) between the ``gas puff'' component and the field of view in most of those configurations. This, in turn, makes it highly desirable that the flux of atoms/molecules from the nozzle be somewhat collimated in the toroidal dimension in order to minimize the loss in spatial resolution due to smearing of field-aligned fluctuation features brought about by chordal integration through the gas cloud. 
    \item The variety of LCFS locations and island locations makes it desirable to have the capability to change the FOV.
    \item Longer pulse/steady-state higher-power operation is planned for W7-X, and plasma radiation from those discharges will result in large heat loads on in-vessel diagnostics with plasma-facing components. These are required to be designed to withstand sustained radiative heat fluxes from the plasma boundary of up to 100 kW/m$^2$. For GPI, this means thoughtful heat load management for its in-vessel components, especially the vacuum window,  plasma-facing components, and nozzle structure.
    \item The heat load concerns precluded the use of in-vessel optical fibers that would transmit light gathered from imaging optics mounted (say) on the vessel wall to the ex-vessel ``fast camera'' component. This required that re-entrant tubes of $\sim$2 m radial extent be used to provide the view of the gas puff and transmit the image to the ``fast camera''.
    
\end{itemize}

In addition to the challenges specific to W7-X, there are challenges that must be overcome for any successful GPI realization, i.e.:
\begin{itemize}
    \item Puff enough gas and gather enough light on turbulence timescales ($\sim1 \mu s$) for good signal-to-noise in the images.
    \item Image with a spatial resolution good enough to discern the turbulence spatial scales of greatest interest. This includes minimizing the toroidal spread of the gas puff at the FOV of the light collection component as well as choosing the spot size of a detector pixel in the object plane of the collecting optics, i.e. in the focal plane in front of the ``gas puff'' component/nozzles.
    \item Do not puff gas in amounts that will significantly perturb either the global plasma or  the $n_e$ and $T_e$ of the region whose turbulence characteristics are being measured. 
\end{itemize}
\subsection{\label{sec:design criteria}Design criteria for Gas-Puff Imaging on W7-X}
The challenges listed in Section \ref{sec:W7-X challenges} formed the basis for the design criteria of the system that was ultimately realized. Those criteria are summarized in Table \ref{tab:design_criteria}, along with the final design decisions. 
\begin{table*}[t]
\caption{\label{design_table}%
Design criteria for the GPI system, the realized design decisions, and their motivations. The ``standard'' W7-X magnetic configuration is the most frequently run n/m = 5/5 island configuration, with n (m) as the toroidal (poloidal) mode number of the island chain. The ``bean'' shape of the plasma is the most vertically extended cross-section of the W7-X plasma.}
\begin{ruledtabular}
\begin{tabular}{cccc}
Parameter&Constraint&
\multicolumn{1}{c}{\textrm{Reason for constraint}}&
\multicolumn{1}{c}{\textrm{Realized value}} \\

\hline
angle between sight&$<15^\circ$&\mbox{ideally parallel}&\mbox{8-11$^\circ$} \\
lines \& local B& &to reduce & \\
 & &spatial smearing& \\
 & & & \\
poloidal/radial&centered btw.& diagnostic should measure& center of FOV is\\
location of FOV&0 and 60 mm out&scrape-off layer fluctuations&$\approx$25 mm outside the\\
 &from LCFS \& contain&\& dynamics at island &LCFS \& contains O-pt. \\
 & island O-pt. of& & \\
 &``standard'' mag. config.& & \\
 & & & \\
radial dist. between& $\leq$150 mm&gas cloud spreads &110 mm\\
FOV \& gas-puff& &toroidally and vertically,& \\
nozzles& &reducing resolution and& \\
 & &brightness with distance& \\
 & & & \\
dist. between puff& $600\leq d \leq 1000$ mm&maximize practical etendue& 778 mm\\
plane and turning& & & \\
mirror& & & \\
& & & \\
toroidal location&within \mbox{15$^\circ$} of&core ITG turbulence&13.8$^\circ$ \\
of FOV&``bean'' shape&predicted to be greatest& \\ 
& &outboard of ``bean'' shape&\\
\end{tabular}
\end{ruledtabular}
\label{tab:design_criteria}
\end{table*}

\section{\label{sec:level1}Realization of GPI on W7-X}
In this Section, we describe the details of the various GPI components as realized on W7-X and provide, in some cases, reasons for the design choices.
\subsection{\label{sec:camera}The fast camera}
\begin{figure}
    \center
    \includegraphics[scale=0.5]{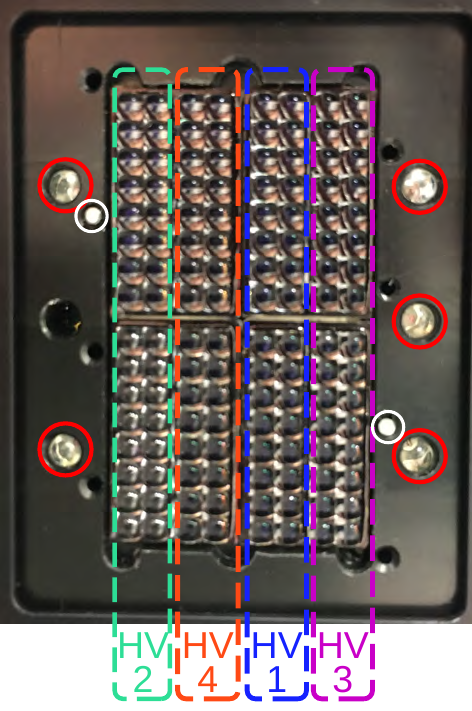}
    \caption{Photo of the APDCAM-10G detector's 8x16 array of APDs covered by the microlenses. The long dashed rectangles indicate pixels connected to the 4 independent high-voltage bias generators. Also shown are the 5 high-intensity red LEDs (circled in red) and the two 1 mm diameter fibers in the detector plane (circled in white).}
   \label{fig:APDCAM detector}
 \end{figure}

In order to maximize the chances of satisfying the crucial criterion of good signal-to-noise in the images with minimal plasma perturbation by the gas puff, we chose to image the collected light directly onto detectors with optimal sensitivity and optimal signal-to-noise at low light incidence. Avalanche photodiodes (APDs) fulfill these sensitivity criteria \cite{Dunai_APDs}. Dynamic range flexibility is provided by changing the high-voltage bias applied to APDs. A camera that utilizes arrays of APDs and provides on-board high-voltage bias control, adequate magnetic field shielding (up to 100 mT), onboard Peltier cooling of the APD arrays, and onboard 10 Gbit optical Ethernet data-transfer capability is commercially available from Fusion Instruments, Kft \cite{FusionInstruments}. The Fusion Instruments APDCAM-10G camera was provided with a custom APD array of 8$\times$16 pixels, utilizing four Hamamatsu S8550 4$\times$8 pixel array chips \cite{HamamatsuS8550} in the 8x16 configuration shown in Figure \ref{fig:APDCAM detector}. All APD pixels are covered by plastic microlenses that increase the effective packing fraction from $\sim$50\% to $\sim$90\%, make the effective pixel size 2.3$\times$2.3 mm, and extend the full detector area to a 20.8$\times$38.3 mm rectangle. The noise on the 14-bit (16384 counts maximum) signals is $\simeq$ 25-30 counts at the controlled detector temperature of 25$^\circ$ C. The camera can be read out to the data archive at frame rates of up to 4 Msamples/s, although 2 Msamples/s has been the typical frame rate.

A $50\times50$ mm-square bandpass interference filter is mounted 14 mm in front of the detector plane. Special high transmission (97-99\%) filters \cite{Chroma} are used (for $H_\alpha$: center-$\lambda$=657 nm with 5.2 nm passband, for He I: center-$\lambda$=587.4 nm with 4.4 nm passband). 

Figure \ref{fig:APDCAM detector} also shows the five high-brightness LEDs and two 1-mm fibers that are located in the detector plane. These are used (during periods of in-vessel access) as light sources for back-illumination through the collection optics onto a target at the toroidal angle of the gas-puffing nozzles that defines the ``FOV registration plane''. The next subsection describes how this back-illumination procedure achieves an accurate registration of the FOV in W7-X vessel coordinates. The two fibers can also be used to couple collected light to spectrometers if visible spectra from the FOV region are desired.   

\subsection{\label{sec:FoV}The field of view}
\begin{figure}
    \centering
    \begin{subfigure}{.5\textwidth}
  \centering
  \includegraphics[width=.95\linewidth]{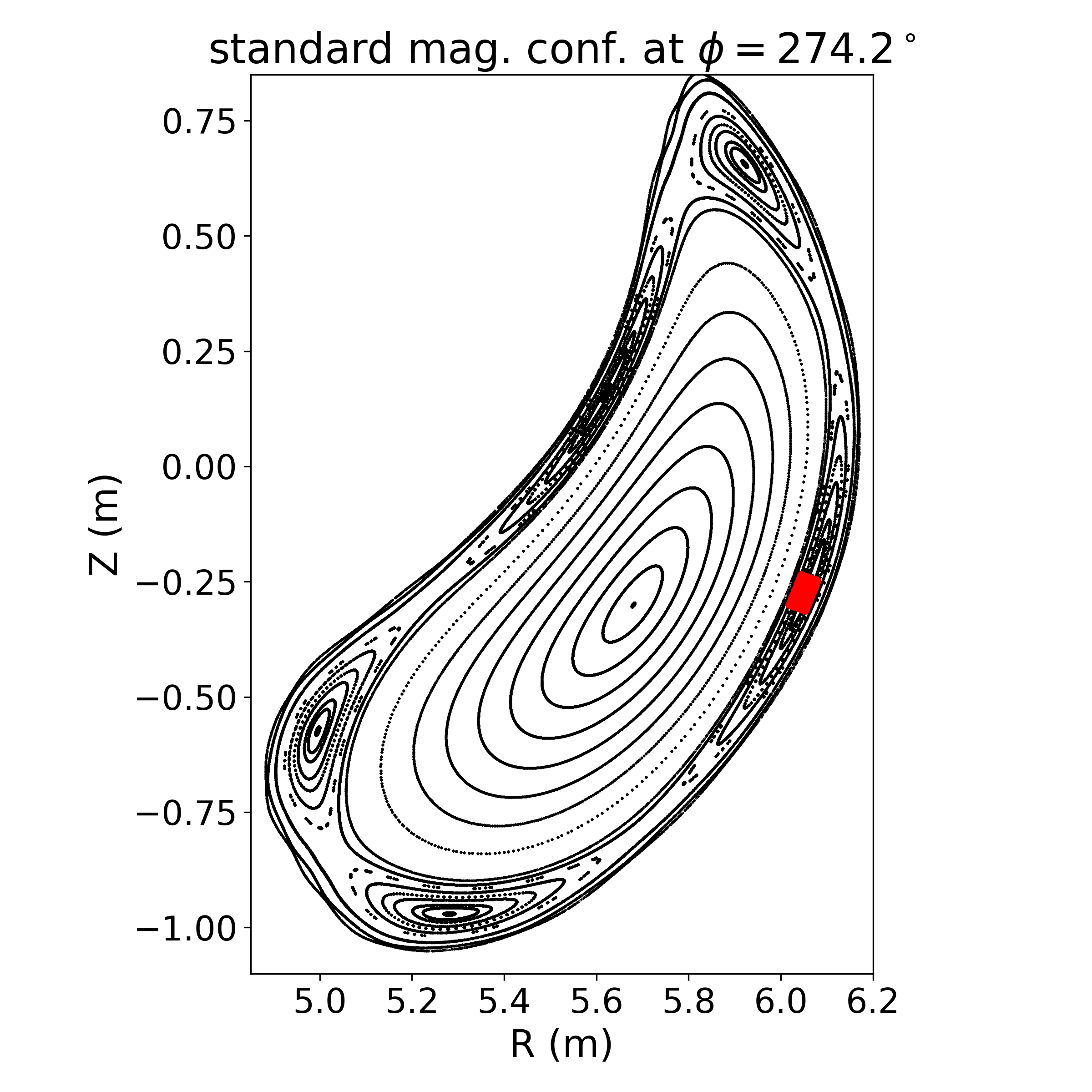}
  \caption{ }
\end{subfigure}%
\begin{subfigure}{.5\textwidth}
  \centering
  \includegraphics[width=.95\linewidth]{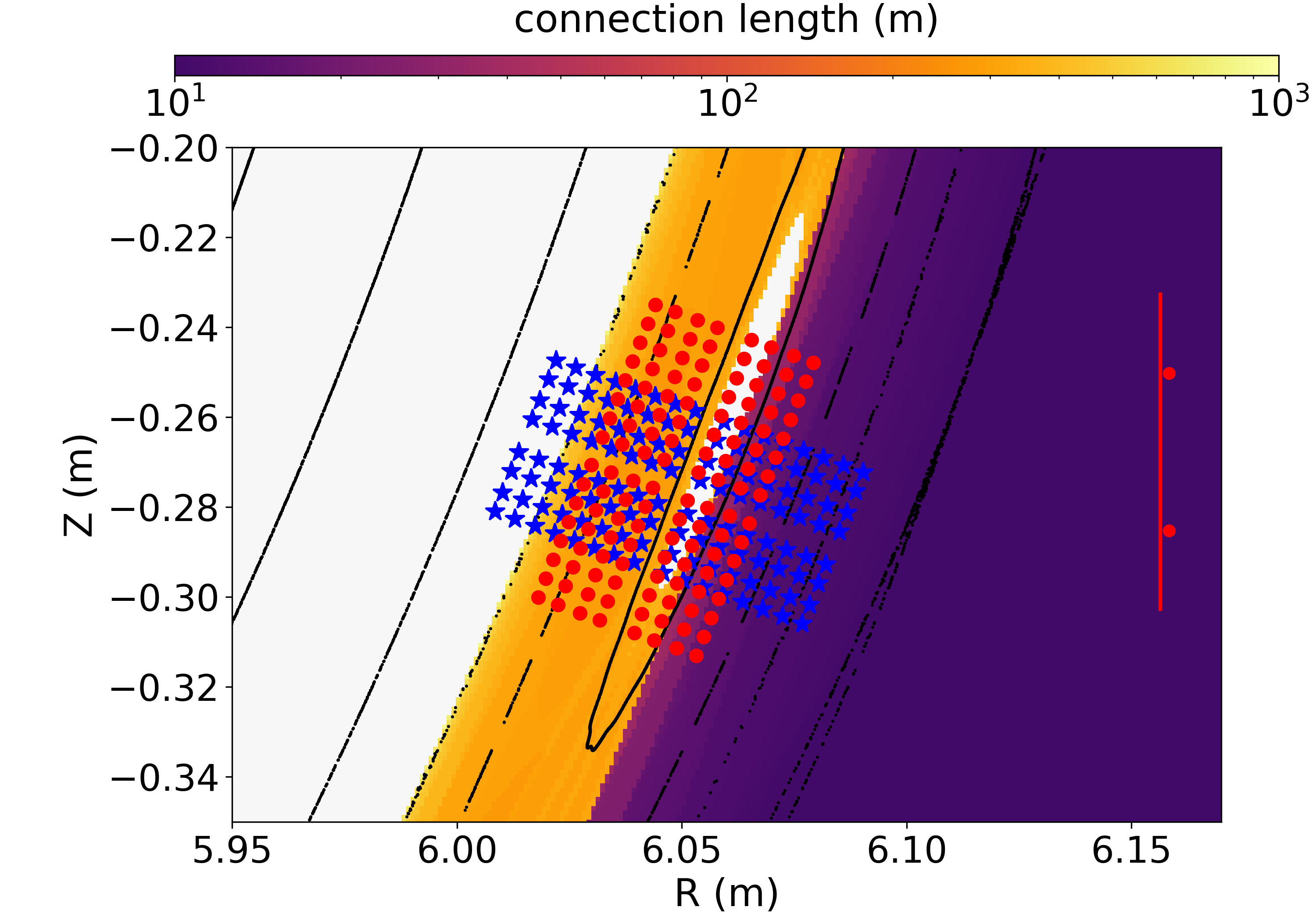}
  \caption{ }
\end{subfigure}%
\caption{(a) Poincar\'e plot showing the structure of the W7-X ``standard'' magnetic field configuration at the toroidal angle of the GPI nozzle and ``FOV registration plane'' of the GPI collection optics. Also shown is the GPI FOV in the ``default'' orientation. (b) Close-up of the GPI FOVs relative to the magnetic island there. View centers in the ``default'' (``rotated'') orientation are shown by the red circles (blue stars). The locations of the nozzles are indicated by the red dots at the right. Also shown color-coded are the  connection lengths in this region.}
   \label{fig:FoV}
\end{figure}

The usefulness of GPI is limited to plasma regions where there is enough atomic line emission from the puffed gas to allow imaging at the appropriate timescales ($\sim1\,\mu s$). In W7-X, this means GPI is limited to the SOL and regions just inside the LCFS, hence the second constraint listed in Table \ref{tab:design_criteria} that the FOV center be within 0 mm and 60 mm radially out from the LCFS. Following the choice of detector in the fast camera (pixel spacing $\simeq$2.3 mm) and a desired spatial resolution ($\sim$5 mm) compatible with the cross-field size scale expected of the SOL filamentary structures ($\sim$10 mm) \cite{Zweben_invited2017}, the demagnification of the collection optics was chosen to be $\sim$0.5 (actual value is 0.510 - see Section \ref{sec:collection optics}). The size of the FOV in the object plane of the collection optics is thus 41 x 75 mm. The toroidal location of the GPI ``FOV registration plane'' is in front of the nozzle. The availability of horizontal ports and mounting choices along with the last three constraints listed in Table \ref{tab:design_criteria} determined the toroidal location of the nozzle at 283.7$^\circ$, i.e. 13.7$^\circ$ clockwise from the ``bean'' cross-section in the 5th of W7-X's five-fold symmetric modules (M5). The cross-section of the ``standard'' magnetic field configuration at that toroidal angle is shown in Figure \ref{fig:FoV}. The FOV-positioning decisions were made in this configuration because it is the one most frequently produced on W7-X. As can be seen in Figure \ref{fig:FoV}, two GPI fields-of-view can be selected for any given experimental run-day by rotating the camera by 90$^\circ$ about the optical axis, a feature built into the camera-mounting apparatus. In the ``default'' FOV orientation, the long dimension is tangential to the local flux surfaces and thus in the poloidal direction, yielding 8 radial columns of pixel views, each with 16 poloidal views. Alternatively, rotation of the fast camera about the optical axis into the ``rotated'' orientation provides 16 radial columns with 8 poloidal views each, covering a greater radial range and including views inside the LCFS. The camera rotations are reproducibly achieved in $\sim$30 minutes of work at the camera location.

\begin{figure}
    \center
    \includegraphics[scale=0.27]{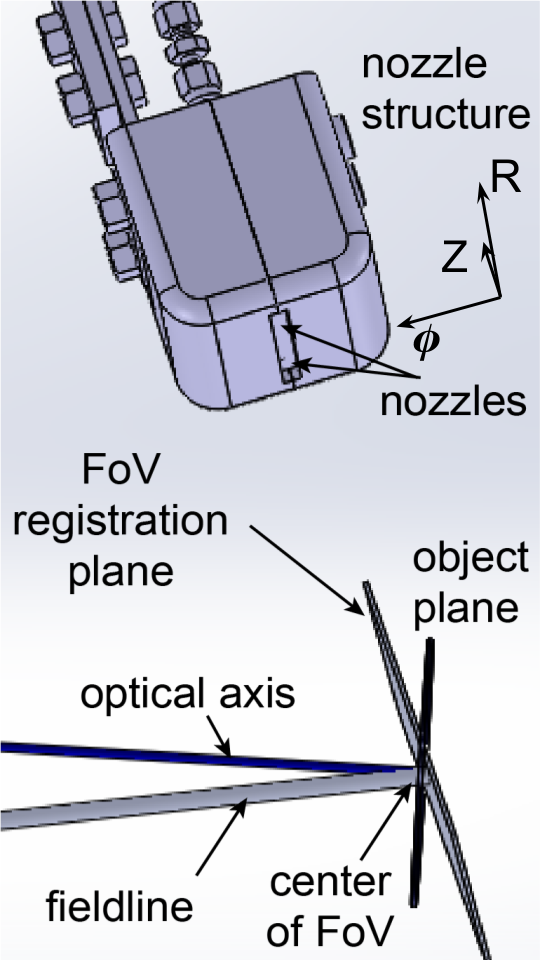}
    \caption{CAD drawing of nozzle structure viewed from above and in front with illustrations of the ``FOV registration plane'', the object plane of the collection optics, the optical axis and a field line, both piercing the center of the FOV. The optical axis is normal to the object plane. The gas cloud issues from the nozzles into the FOV.}\label{fig:viewing_geometry}
 \end{figure}

The ``default'' FOV, shown in red in Figure \ref{fig:FoV}, includes an island O-point in the ``standard'' configuration (one of the design criteria) and is entirely in the SOL. The choice of its radial location was also informed by (time-independent) modeling of emission resulting from the interaction of the SOL plasma with the gas puff (see Section \ref{sec:modeling}). The FOV was placed so that the radial location of the peak line emission predicted by the modeling was just inside the outer edge of the FOV under ``typical'' SOL conditions. Accurate registration of the FOV is achieved by mounting a target onto the nozzle structure whose location is known in W7-X vessel coordinates from in-vessel metrology. This target defines the ``FOV registration plane'' which is orthogonal to the front face of the nozzle structure and is essentially in the $\phi=274^\circ$ (R,Z) plane. The geometry is illustrated in Figure \ref{fig:viewing_geometry}. The fast-camera LEDs and fibers (see Figure \ref{fig:APDCAM detector}) are back-illuminated through the collection optics and imaged onto the target by the optics. Since the pixel positions relative to the LEDs/fibers are known, the locations viewed by the camera pixels are accurately registered in W7-X coordinates, thereby defining the FOV in this plane. In the ``standard'' configuration, the radial/normal-to-flux-surface extent of the ``default'' FOV is from 5 mm to 44 mm outside of the LCFS. For this and the other common W7-X magnetic configurations, the FOV (relative to the LCFS) are listed in Table \ref{tab:FoV_extents}, where it can be seen that there is good coverage in all but the ``high iota'' and ``high mirror'' configurations, for which the emission in the outer views is weak because they are so far out in the SOL.

\begin{table}[!htb]
\caption{Radial extents of the GPI FOV in the ``default'' and ``rotated'' camera orientations, calculated assuming plasma $\beta=0$ and no toroidal plasma current.} \label{tab:FoV_extents}
\setlength\tabcolsep{0pt} 

\smallskip 
\begin{tabular*}{\columnwidth}{@{\extracolsep{\fill}}ccc}
\toprule
Configuration&Radial coverage relative to LCFS [mm]&Comment\\
\hline
standard&\mbox{5 to 44 (``default'')}&includes O-pt \\
&\mbox{-12 to 65 (``rotated'')}&includes O-pt \& LCFS\\
\hline
low iota&\mbox{-18 to 21 (``default'')}&spans LCFS \& includes X-pt\\
&\mbox{-35 to 42 (``rotated'')}&spans LCFS \& includes X-pt\\
\hline
high iota&\mbox{31 to 70 (``default'')}&weak emiss. on outboard views\\
&\mbox{13 to 90 (``rotated'')}&\\
\hline
low mirror&\mbox{-3 to 36 (``default'')}&spans LCFS \\
&\mbox{-20 to 56 (``rotated'')}&spans LCFS\\
\hline
high mirror&\mbox{31 to 71 (``default'')}&includes O-pt \\
&\mbox{14 to 91 (``rotated'')}&includes O-pt\\
\bottomrule
\end{tabular*}
\end{table}

\subsection{\label{sec:mirror_shutter}The turning mirror, shutter, and vacuum window}
The gas puff emission is viewed from the horizontal port adjacent to the one holding the nozzle structure (Figure \ref{fig:GPI_CAD}). The views are designed to be as close to B-field-aligned as is feasible. We set as a constraint (the first in Table \ref{tab:design_criteria}) that the sight lines be within 15$^\circ$ of the field lines local to the gas puff. A turning mirror is required to fulfill this requirement, which we decided to combine with the need to shutter the vacuum window of the re-entrant tube holding the collection optics. We mounted a polished stainless steel mirror (flatness of 1 $\lambda$ at 630 nm and 60/40 scratch dig) on the back side of a hinged shutter plate and control the opening/closing of the shutter and mirror with a pneumatically actuated linear motion vacuum feed-through. The full-open angle of the shutter sets the location of the FOV, so it is critical that hard stops provide a very reproducible ``open'' condition. Repeated testing showed that the views in the ``FOV registration plane'' were reproducible to within $\leq\pm$1 mm.

The desire to field-align the sight lines needed to be balanced with the need to minimize/eliminate interaction between the shutter and the plasma, including fast ions from the neutral beams. This compromise resulted in the following:
1) we placed the center of the open turning mirror at W7-X coordinates $[R,\phi,Z]=[6363\,\mathrm{mm},4.900\,\mathrm{rad},-210\,\mathrm{mm}]$; 2) angles between the sight lines and the B-field local to the gas puff are therefore from 8 to 11$^\circ$; 3) since we were free to choose the vertical location of the turning mirror, the sight lines and the local field lines piercing the ``FOV registration plane'' are at essentially the same angle ($\pm2^\circ$) relative to horizontal planes; 4) the shutter is fully recessed into the port extension when closed; and 5) the part of the open shutter closest to the plasma is $\sim$115 mm from the LCFS of the largest of the most common magnetic configurations (i.e. the ``low-iota'' configuration). Reciprocating probe measurements have shown that the non-radiative power fluxes at locations $>$100 mm outside the LCFS surface are negligible \cite{drews_MPM_profiles_NF_2017, Killer_NF_2019}. Nonetheless, radiative heat fluxes on the plasma-facing shutter of up to 100 kW/m$^2$ had to be designed for, an issue discussed in Section \ref{sec:heatload}.

The 98 mm clear-aperture vacuum window in the re-entrant tube is a critical component for machine safety. The two main threats to the window integrity are radiative heating by the plasma and stray ECRH power. These risks were mitigated by recessing the window by $\sim$660 mm up the water-cooled re-entrant tube, whose inner diameter is 100 mm up to that window location. The fused silica window has a broadband anti-reflection coating and is rated for temperatures up to 200$^\circ$C and for heating rates up to 25$^\circ$C/min. \cite{Vacom}.  

\subsection{\label{sec:collection optics}The collection optics}

\begin{figure*}
    \center
    \includegraphics[width=0.95\textwidth]{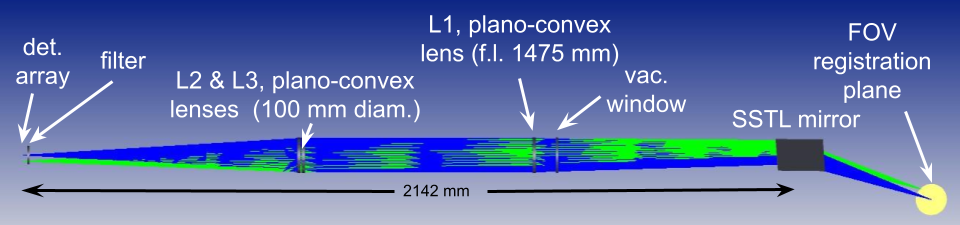}
    \caption{Optical layout of collection optics.}
   \label{fig:Zemax}
 \end{figure*}

In addition to the turning mirror, three lenses in the re-entrant tube on the atmosphere side of the vacuum window make up the ``collection optics''. They collect light from sight lines passing through the region in front of the nozzles, transmit it up the re-entrant tube and focus it onto the detectors. It was a primary design priority to collect as much light as was feasible, given the constraints of the size and availability of windows and lenses. This is accomplished using three 100 mm diameter plano-convex glass lenses with broadband anti-reflection coatings on all surfaces.
The optical design is shown in Figure \ref{fig:Zemax} and is summarized here: the front lens, L1, is located just behind the vacuum window and renders light from the object plane parallel so that it passes down the tube. As illustrated in Figure \ref{fig:viewing_geometry}, the object plane intersects the vertical ``FOV registration plane'' (normal to the nozzle face) at their respective center points. The ``FOV registration plane'' is essentially the $\phi$=274$^\circ$ (R,Z) plane, while the object plane is rotated by 20$^\circ$ about a vertical axis and by 4.2$^\circ$ about a horizontal axis with respect to that plane. In other words, the optical axis strikes the ``FOV registration plane'' at a downward angle of 4.2 deg, while the total angle between the optical axis and the surface normal of this plane is 20$^\circ$. The focal length of L1 is 1475 mm, equal to the distance from L1 to the ``FOV registration plane'' along the optical axis. Lenses L2 \& L3 are placed together $\sim$665 mm beyond L1 and focus the parallel light onto the detector plane with an effective focal length of 753 mm, yielding a demagnification of 0.510. The 20$^\circ$ angle between the “object plane” and the “FoV registration plane” results in some differences in the optical properties across the FOV. The (geometrical) optics imaging properties of the system have been obtained using the ZEMAX-OpticStudio \cite{Zemax} commercial software. The modeled images of points in the detector plane back-imaged into the “FoV registration plane” have diameters (90\% enclosed energy) ranging from $\approx 0.08$ mm to $\approx$0.3 mm, with the largest spot sizes occurring at largest and smallest $R_{maj}$ locations in the FOV. However, this imaging only increases the FWHM of back-imaged pixels from a perfectly-imaged 4.5 mm to 4.6 mm (minimum) and 4.7 mm (maximum) respectively. Thus, we will treat the optical resolution in the ``FoV registration plane'' as $\approx4.65$ mm. Thus 2.3 × 2.3 mm pixels collect essentially all of their incident light from 4.65 × 4.65 mm areas in the ``FoV registration plane''. The aperture stop in the system is the 96 mm diameter clear aperture of the L1 lens mount. The etendue for an on-axis pixel is $6.5\times10^{-2}$ mm$^2$ ster. Light from off-axis points is vignetted because it is at a small angle with respect to the central ray on being rendered parallel by L1, yielding an etendue for the most off-axis pixels (the 4 corner pixels) of $4.7\times10^{-2}$ mm$^2$ ster.
The lenses are secured in the re-entrant tube using a removable stiff frame of angled bars that is secured at each end. This can be seen in the cut-away rendering in Figure \ref{fig:GPI_CAD}. 

\subsection{\label{sec:nozzle}The gas puffing component}

The gas puffing component of the system consists of the nozzles, the nozzles' housing, a capillary feed line, a gas vacuum feed-through, an injection valve, and a gas control system. The first four of these items are shown in Figure \ref{fig:GPI_CAD}. The gas control system (not shown in Figure \ref{fig:GPI_CAD}) is mounted on a panel and connects to the injection valve through a $\sim$1 m long flexible stainless steel (SSTL) hose. 

We wanted to locate the nozzle structure as close to the plasma as feasible while avoiding direct heating by plasma conduction and convection. As noted in the previous subsection, reciprocating probe measurements showed that conducted/convected power was quite small at locations $\gtrapprox$ 100 mm into the outboard SOL. Considering the range of LCFS locations of the available W7-X magnetic configurations, we placed the front face of the nozzle housing $\approx$95 mm radially outward from the LCFS of the outermost of the more common configurations. This places the nozzles, which are recessed 2 mm into the housing, $\approx$135 mm from the LCFS in the ``standard'' configuration and $\approx$110 mm from the FOV center. The vertical center of the nozzles was set to align with the FOV center.
Its toroidal location was set by mounting it onto the water-cooled port protection liner, thereby minimally affecting viewing access by other diagnostics utilizing that port. The in-vessel metrology activity mentioned in Section \ref{sec:FoV} registered the location of the nozzle housing front face in W7-X plasma vessel coordinates at $[R,\phi,Z]=[6156.4\,\mathrm{mm},4.7856\,\mathrm{rad},-267.7\,\mathrm{mm}]$.

\begin{figure}
    \center
    \includegraphics[scale=0.4]{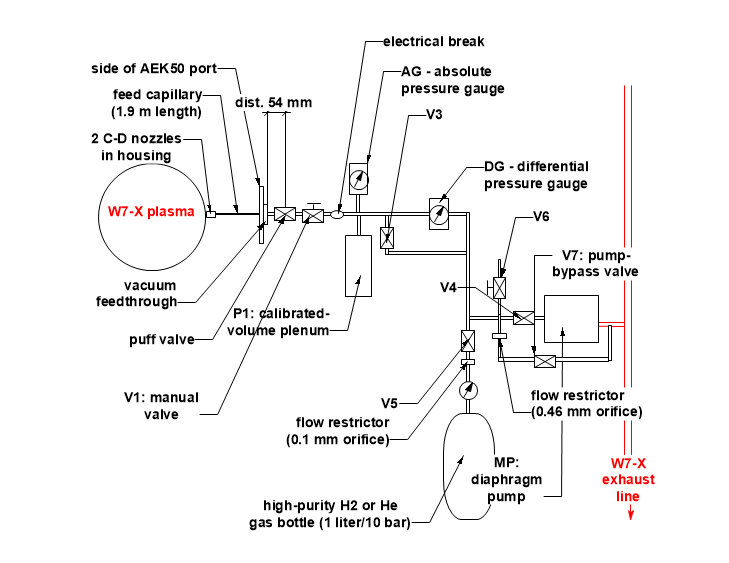}
    \caption{Schematic of GPI gas control system.}
   \label{fig:gas_control_system}
 \end{figure}

We surrounded the SSTL nozzle body in a graphite housing made from the same material that is used for the W7-X divertor targets. A short section of 1 mm I.D. capillary that couples to the nozzle body via a press fit exits radially out the back of the housing and connects with a 1.9 m length of 1 mm capillary that runs along the side of the port liner to a vacuum feedthrough mounted on the port extension close to the port flange. The puff valve is mounted as close as possible to the feedthrough as part of the effort to minimize the volume between the nozzles and the puff valve since that volume of gas ($V_{min}$=3.02 ml) must necessarily flow into W7-X. Thus the \textit{minimum} gas load on W7-X is $p_o\times V_{min}$, where $p_o$ is the plenum pressure backing the puff valve. The gas control system performs several functions in addition to opening the puff valve at programmed times and durations, which it can do up to four times during a single W7-X shot/program. The gas control hardware consists of a plenum volume, an absolute pressure gauge to measure the pressure in the plenum, a differential pressure gauge to measure the rate of change in the pressure occurring during each gas puff, valves to access or isolate various parts of the system, and a mechanical pump to evacuate the system into the W7-X exhaust line, when necessary. A schematic of the GPI gas control system is presented in Figure \ref{fig:gas_control_system}. 
The timing for the opening and closing of the valves, as well as the monitoring and
digitizing of the system pressures, is done by a programmed Red Pitaya's FPGAs/CPU/digitizers. It is mounted on the gas control panel and is interacted with using a Python graphical user interface program running on a network computer.

The plenum pressure is manipulated by the valves V5 and V4 connecting it with either the regulated gas bottle or the mechanical pump while feeding back on the pressure of the absolute pressure gauge, AG \cite{MKS_121A}. Measurements of the gas puff flow rate and total amount puffed are accomplished by digitizing the differential pressure from gauge DG \cite{MKS_221B} after closing the normally-open valve V3 for the duration of the gas puffs. With V3 closed, the time history of the pressure difference between the pre-puff pressure and the plenum pressure is measured, and the time derivative of that differential pressure times the plenum volume ($V_{plenum}$=0.802 l) provides the gas flow rate out of the plenum. The pressure range of the differential gauge is 0 to 39 mbar, with an estimated error of $<$1\% for typical pressures. Thus the maximum measureable amount of gas input into a single W7-X plasma is limited to 31 mbar\,l, which is greater than the typical amount delivered in four puffs, $\sim$12 mbar\,l or $\sim$3 mbar\,l per puff. 

\subsubsection{\label{sec:nozzle design}Nozzle design and gas puff flow rate measurements}
Having defined and registered the locations of the FOV and the nozzles, the radial distance between the nozzles and the center of the FOV is 110 mm, satisfying one of the design criteria listed in Table \ref{tab:design_criteria}. However, that large distance made improved collimation of the gas cloud a high priority. Another factor in choosing the nozzle design was the requirement of minimizing the perturbation of the plasma by the gas puff, which is essentially a constraint on the gas flow rate. In this subsection, we will present 1) the details of the ``converging-diverging'' (C-D) nozzle design, which was driven by the collimation and perturbation constraints, 2) its fabrication, and 3) the measured flow rates from the nozzles and comparison to predictions for C-D nozzles.

We characterize the degree of collimation using the half angle of the gas cloud cone as it expands from a single nozzle into the relative vacuum, defined here as $\alpha_{1/2}=$ tan$^{-1}($HWHM$/\Delta)$, where HWHM is the half-width at half-maximum of the gas pressure distribution in a normal plane at a distance $\Delta$ from the nozzle. The types of nozzles we considered were: unshaped 1 mm diameter capillary tubes, de Laval nozzles providing, in theory, optimal collimation \cite{Vlad_nozzle}, and C-D nozzles \cite{Collis_2006} with good collimation. The de Laval shape was too difficult to fabricate at the size estimated for the desired flow rate. C-D nozzles were difficult, but possible, to fabricate. Unshaped capillaries were a last resort since our only guidance was a measurement from Ref. \cite{Griener_piezo} of $\alpha_{1/2}\approx$25$^\circ$ for D$_2$ at 600 mbar backing pressure, and we desired better collimation than that. 

We chose to use two nozzles, each with a C-D shape, separated vertically by 35 mm. The two nozzles with that separation were necessary in order to properly ``illuminate'' the full poloidal FOV in the ``default'' camera orientation.
For the C-D nozzle, one specifies the converging angle of the nozzle shape on the high gas pressure side, the diverging angle on the outflow side, and crucially the ``throat'' diameter where the cones meet. Under isentropic conditions and with a high enough pressure difference between the input and output sides, Mach 1 flow is achieved at the throat with a flow rate there that is given by \cite{anderson1990modern}
\begin{equation}
    \dot{N_p}=n^{throat} v^{throat} A^{throat}={\frac{{C (\gamma) p_o}}{kT_o}} \sqrt{\gamma R T_o} A^{throat}
    \label{eq:CD_flow_rate}
\end{equation}
where $\dot{N_p}$ is the gas flow rate [\#/s], $n^{throat}$ is the gas particle density (at the throat), $v^{throat}$ is the Mach 1 speed, $A^{throat}$ is the throat area, $C(\gamma)$=0.58 for H$_2$ (and =0.57 for He) is a constant that relates the gas density and temperature at the throat to the gas density and temperature of the backing reservoir and depends only on $\gamma$, the ratio of specific heats for the specific gas (1.41 for H$_2$ and 1.67 for He), $p_o$ is the backing reservoir pressure in Pa, $T_o$ is the reservoir gas temperature in K, $k$ is Boltzmann’s constant, and $R$ is the ideal gas constant for that gas, e.g. 4124 J/kg/K for H$_2$. 

Measurements of $\alpha_{1/2}$ on a C-D nozzle with a 350 $\mathrm{\mu}$m throat diameter \cite{Collis_2006} showed that $\alpha_{1/2}$ decreased significantly as the backing pressure increased over the range of interest ($\sim$0.4 to $\sim$1.3 bar). Since the flow rate is proportional to the backing pressure, we had to balance the improved collimation at the higher backing pressures with the risk of perturbing the W7-X plasma. According to Eq. \ref{eq:CD_flow_rate}, the throat diameter of the C-D nozzle design and the backing pressure determine the gas flow rate. The maximum acceptable gas flow rates were determined by modeling a hypothetical gas puff that perturbs the W7-X global plasma density by 10\%.
This modeling therefore informs the choice of throat diameter for the C-D nozzle shape. 

We modeled the total electron inventory in the W7-X plasma, $N_e^{tot}$, as being sustained by a recycling source, $\Phi_{rec}$, that is proportional to the total inventory $N_e^{tot}$ and balanced by a sink defined by an effective particle confinement time, $\tau_p$. We defined another source, $\Phi_{ext}(t)$, with time-dependence similar to that predicted from finite-element analysis of gas transport occurring during a 50 ms opening of the puff valve to our plenum and accounting for flow through the 1.9 m long feed capillary to two C-D nozzles with throat diameters of roughly 100 $\mu$m. The time-profile of the external source was a $\sim$ 40 ms rise time, a 50 ms flattop, and a 0.5 s exponential decay. This external puff was assumed to fuel the plasma with an efficiency, $\epsilon_{fuel}$, of 0.8, as will be discussed below. The model equation was thus:
\begin{equation}
    \frac{N_e^{tot}}{dt}=\Phi_{rec} + \epsilon_{fuel} 2  \Phi_{ext}(t) - \frac{N_e^{tot}}{\tau_p}
    \label{eq:inventory_model}
\end{equation}
$\tau_p$ was taken to be 300 ms (using Fig. 8 from Ref. \cite{Schmitz_2021}). Taking $N_e^{tot}=V_{plasma} C n_{el}$, where $V_{plasma}$ is the plasma volume ($\approx30$ l), $n_{el}$ is the line-integrated density, and $C$ is the ratio between the volume-averaged and line-integrated density ($\approx$0.8 $m^{-1}$), $\Phi_{rec}$ is evaluated using the measured $n_{el}$ with $\Phi_{ext}=0$. We note that using essentially the same model and actual W7-X discharges (produced in 2018) into which He gas was puffed using a divertor gas puff system, Ref. \cite{Kremeyer_2022} evaluated $\tau_p$ as (0.258$\pm$0.124 s) and the fueling efficiency of those divertor puffs to be in the range of 0.3 to 0.44. Using other 2018 discharges with $n_{el}$ values of $2.2\times10^{19}$, $5.5\times10^{19}$, and $9\times10^{19}$  m$^{-2}$, we modeled the response to an external gas puff with the time-profile described above and the conservative guess for $\epsilon_{fuel}$ of 0.8. 
The peak flow rates that yielded a 10\% increase in the $n_{el}$'s of those three modeled discharges are listed in Table \ref{tab:perturbation_study}. Also shown there are the backing pressures that according to Eq. \ref{eq:CD_flow_rate} would provide those rates through two C-D nozzles, each with a 70 $\mathrm{\mu}$m throat diameter. A 70 $\mathrm{\mu}$m throat diameter was the smallest that could reasonably be fabricated.

\begin{table}[!htb]
\caption{Peak H$_2$ flow rates yielding a modeled 10\% density perturbation, and corresponding required backing pressures for two 70 $\mathrm{\mu}$m throat diameter C-D nozzles.}
\label{tab:perturbation_study}
\setlength\tabcolsep{0pt} 

\smallskip 
\begin{tabular*}{\columnwidth}{@{\extracolsep{\fill}}ccc}
\toprule
W7-X $n_{el}$&peak flow rate [\#/s]&backing pressure [bar]\\
&(10\% perturbation)&\\
\hline
$2.2\times10^{19}$ m$^{-2}$&$4\times10^{19}$&0.34\\
\hline
$5.5\times10^{19}$ m$^{-2}$&$14\times10^{19}$&1.17\\
\hline
$9\times10^{19}$ m$^{-2}$&$30\times10^{19}$&2.5\\
\bottomrule
\end{tabular*}
\label{tab:perturbation_flowrates}
\end{table}

Thus, we specified a 70 $\mu$m throat diameter. The half-angle for the diverging cone was specified as 25$^\circ$, since that was seen to be in the optimal range in Ref. \cite{louisos2008design}. A 45$^\circ$ half-angle was chosen for the converging cone. The two nozzles were cut into a trough milled down to 1 mm  thickness in a SSTL blank with an overall thickness of 3 mm.

\begin{figure}
    \center
    \includegraphics[scale=0.4]{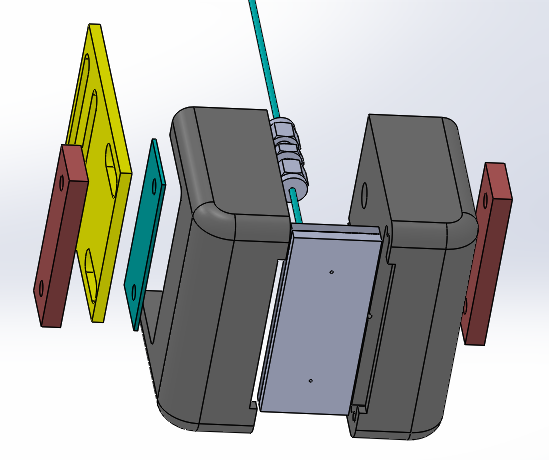}
    \caption{CAD rendering of the exploded nozzle structure. From left to right: left clamp, bracket for mounting to the port extension fixture, Sigraflex pad \cite{Sigraflex} for better heat conduction, left half of the graphite housing, the welded SSTL piece with two C-D nozzles (separated by 35 mm vertically) and feed capillary, right half of the graphite housing, and right clamp. The short feed capillary exiting the back of the nozzle structure connects to the longer 1.9 m feed capillary. The clamping and mounting bolts are not shown.}
   \label{fig:exploded_nozzle}
 \end{figure}

Two vendors willing to attempt fabrication to those specifications were identified. One vendor used fine drills with tips cut at the desired angles; the other used a laser to sculpt the shape by varying the focal spot. Both fabrications were to be electro-polished to a surface finish with a Roughness Average $Ra$ better than 0.4 $\mu$m. Examination of results with a microscope showed that the mechanical drilling produced far superior results. The cone walls were smooth, and the throat was well-defined, albeit with a diameter of $\approx83 \mathrm{\mu}$m, larger than the specification. The walls of laser-drilled nozzles were rippled; the throat hole was ragged, and this fabrication was not used. The nozzle plate was then vacuum welded onto a back cover plate that holds a press-fit length of 1 mm I.D. capillary protruding out of the backside of the back plate. The assembly was vacuum leak-tested by pumping on the capillary after temporarily blocking the nozzle holes. An exploded view of the nozzle structure is shown in Figure \ref{fig:exploded_nozzle} and an assembled rendering is shown in Figure \ref{fig:viewing_geometry}. 
\textbf{\begin{figure}[hbt!]
    \center
    \includegraphics[scale=0.5]{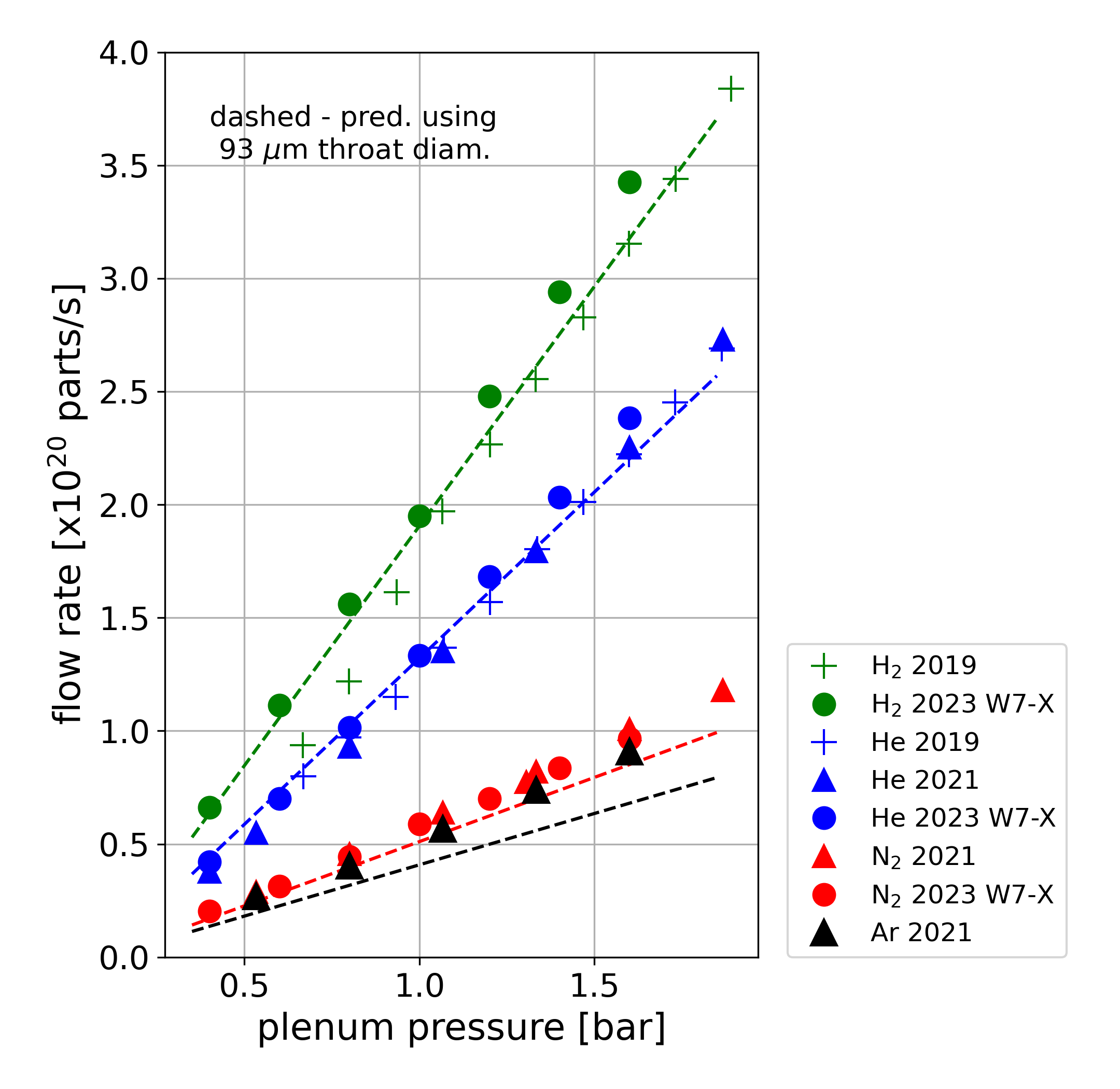}
    \caption{Flow rate measurements of four gases as a function of the measured plenum pressure. The dashed lines are the calculated flow rates using Eq.\ref{eq:CD_flow_rate} and assuming a 93 $\mu$m diameter throat for each of the two nozzles. The solid circles are from puffs into W7-X, while the $+$'s and $\triangle$'s are from lab measurements.}
   \label{fig:flowrates_vs_pressure}
 \end{figure}}

The flow rates of H$_2$ and He, the gases of interest for GPI, as well as N$_2$ and Ar were measured multiple times over 4 years using the gas control system. The 1.9 m length of feed capillary was present in the system for these measurements. The results are shown in Figure \ref{fig:flowrates_vs_pressure}.
For one set of H$_2$, He, and N$_2$ measurements, the puffs were into the W7-X vacuum vessel with the gate valves to the pumps closed and with the vessel pressure measured by a calibrated ASDEX-Upgrade Baratron gauge. Comparisons of the total amount of gas puffed during each sequence at each plenum pressure for each gas showed that $(\Delta p_{GPI}  \times  V_{plenum})$ and $(\Delta p_{W7-X}  \times  V_{W7-X})$ were the same to within 5\%. This provided independent confirmation of the accuracy of the GPI differential pressure measurements. Regarding the flow rate measurements, we note the following:
\begin{itemize}
\item The flow rates are linear with backing pressure as predicted by Eq. \ref{eq:CD_flow_rate}.
\item The flow rates are $\approx$25\% larger than what is predicted by Eq. \ref{eq:CD_flow_rate} using the measured throat diameter. We do not know the reason(s) for this. Note that, in the figure, the 93 $\mu$m throat diameter was chosen in order to provide a good match to the experimental flow rates.
\item The scaling among the 4 different gases is approximately what is predicted by Eq. \ref{eq:CD_flow_rate}. It is good for H$_2$ and He, but somewhat worse for N$_2$ and Ar relative to H$_2$ and He.
\end{itemize}

\subsubsection{\label{sec:gas cloud measurements}Measurements of gas cloud extent}

As noted in the preceding subsection, we had reason to believe that C-D shaped nozzles would provide some degree of collimation of the outflowing gas, and that it should improve with increasing plenum pressure \cite{Collis_2006}. We chose to measure this for the nozzles as fabricated.
In this subsection, we present the measurements of the spatial distributions of the gas emerging from the nozzles. These distributions were measured using the vacuum chamber and the computer-controlled positioning system capable of scanning a probe in a 2D plane of the VINETA II linear plasma device \cite{VINETA_RSI} at IPP. By mounting a custom miniaturized hot filament ionization gauge onto VINETA's positioning system, we measured the spatial distributions of the local pressure within the puffed gas cloud as functions of gas backing pressure.

\begin{figure}
    \center
    \includegraphics[scale=0.5]
    {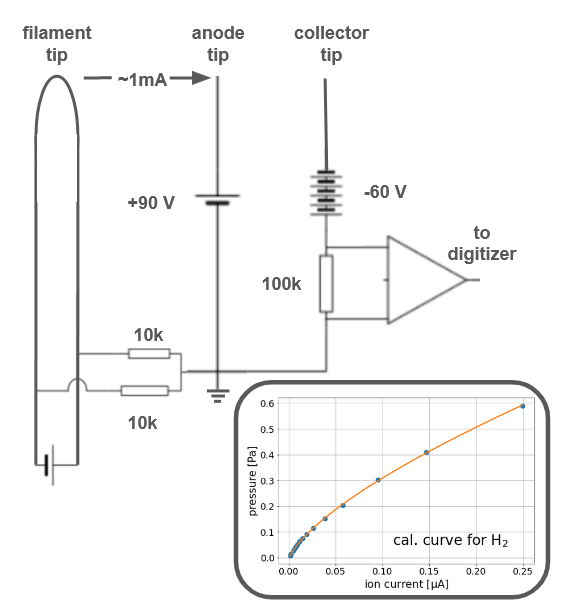}
    \caption{Circuit diagram for the miniaturized ion gauge probe. Inset: an example of the calibration of the ion gauge for H$_2$ gas}
   \label{fig:ion_gauge circuit}
 \end{figure}

The vacuum chamber of VINETA II is roughly cylindrical, $\sim$1.6 m long and $\sim$0.97 m in diameter, with a typical base pressure of $\sim2\times10^{-4}$ Pa. The custom ionization gauge is based on a design from Ref. \cite{timo_thesis}. It consists of three tips protruding $\sim$3 mm from channels in a long 5 mm diameter ceramic cylinder. One tip is a U-shaped loop of tungsten wire that is heated to thermionic emission and acts as an electron source (the cathode at V=0, i.e., ground) for a second probe (the anode) that is biased (at +90 V) to draw $\sim$1 mA of electron current. The third probe (the collector) is biased (at -60 V) to collect ions of the gas local to the probe tips that have been ionized by the cathode-to-anode electrons. The anode and collector are separated by $\approx$1 mm. The collector current is amplified and digitized. The electrical circuit is shown in Figure \ref{fig:ion_gauge circuit}. 
 
 The measured ion currents are small (0 to 0.25 $\mu$A), non-linear (but monotonic) functions of the gas density at the probe, and dependent upon the gas species. Calibrations of the ion current response to the pressure of each puffed-gas species were performed by introducing increasing amounts of gas into VINETA II via a needle valve and logging the absolute pressure measured by a Baratron capacitive pressure gauge along with the measured ion current. The calibration range covered the pressure range encountered in the measurements on the GPI puffs into VINETA II (from about 0.005 to 0.5 Pa), and calibrations were performed for each gas species and after each vacuum break. An example of a calibration curve for H$_2$ is shown in the inset of Figure \ref{fig:ion_gauge circuit}.

\begin{figure}
    \center  
    \includegraphics[scale=0.5]{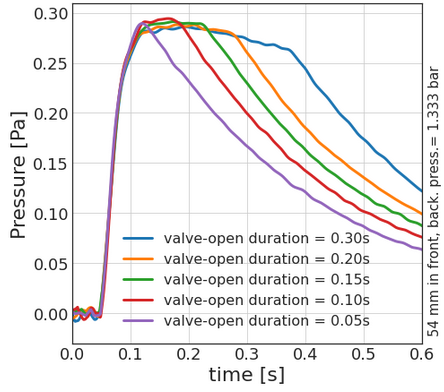}
    \caption{Time histories of pressure in front of the top nozzle for different valve-open durations. The backing pressure was 1.333 bar of He gas.}
   \label{fig:time_histories_vs_valve_duration}
 \end{figure}
 
The miniaturized ion gauge probe was scanned spatially within planes that were 54 and 102 mm in front of and normal to the 2-nozzle plate. Additionally, time histories at a single location 54 mm directly in front of the top nozzle for puffs with different valve-open durations were measured, shown in Figure \ref{fig:time_histories_vs_valve_duration}. The rise time is $\approx$ 40 ms, and the decay time is $\approx$ 350 ms. The long decay time is consistent with our finite-element modeling of the gas transport through the system, and indicative of the gas in the volume between the valve and the nozzles ($V_{min}$) draining out of the nozzles after the valve is closed. Also evident in Figure \ref{fig:time_histories_vs_valve_duration} are the ``flattop'' durations that roughly match the valve-open durations $\geq0.1$ s.

\begin{figure}
   \includegraphics[scale=0.4,center]{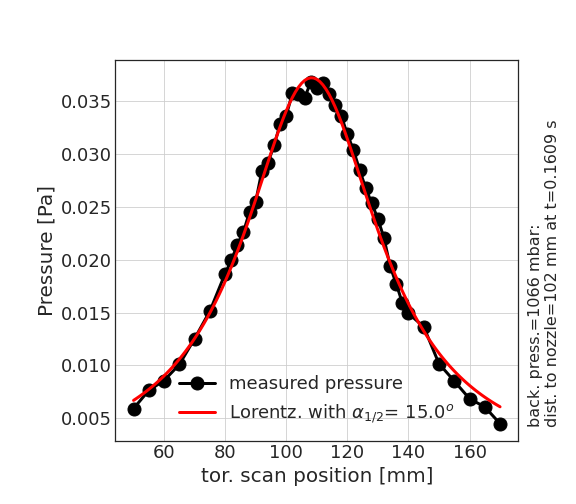}
    \caption{Measured gas puff pressure profile in the toroidal direction 102 mm in front of the nozzles for a plenum pressure of 1.07 bar.}   \label{fig:toriodal_cloud_profile}
 \end{figure}

\begin{figure}
   \includegraphics[scale=0.4,center]{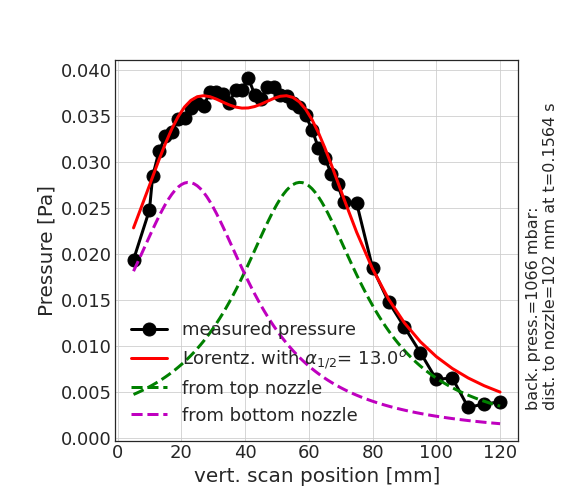}
    \caption{Measured gas puff pressure profile measured in the Z direction 102 mm in front of the nozzles for a plenum pressure of 1.07 bar.}   \label{fig:vertical_cloud_profile}
 \end{figure}

\begin{figure}
    \center
   \includegraphics[scale=0.36]{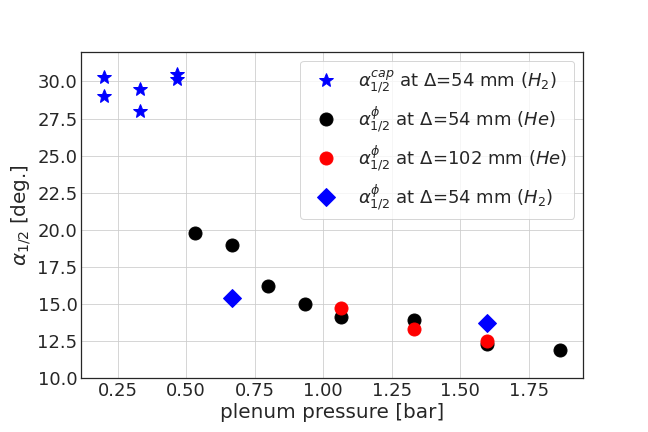}
    \caption{Summary of the gas puff's toroidal half-angle vs. plenum pressure measurements. The black and red circles are for He puffs measured in the 54 mm and 102 mm planes, respectively. The blue diamonds are for H$_2$ puffs. The blue stars are $\alpha_{1/2}$ for H$_2$ distributions from a single unshaped 1 mm I.D. capillary and are included for comparison with those from the C-D nozzles. }   \label{fig:half_angle_vs_plenum_pressure}
 \end{figure}

2D time-resolved pressure measurements were made for He puffs at eight backing pressures in the 54 mm plane and at 3 backing pressures in the 102 mm plane, and for H$_2$ puffs at two backing pressures in the 54 mm plane. For each gas puff, the probe responded to the rapid direct pressure rise within the gas cloud as well as the less swift ``base pressure'' rise in VINETA II. The latter was therefore measured with the probe well away from the nozzle during the puff and subtracted from the overall signal at each measurement location within the gas cloud.  Full 2D scans showed left-right (toroidal) and up-down (vertical) symmetry of the gas distribution around the center of the nozzle structure. An example of a pressure distribution along a cut in the toroidal dimension at the height of the top nozzle is shown in Figure \ref{fig:toriodal_cloud_profile}. 
The distribution is well described by a Lorentzian whose HWHM is equal to $\alpha_{1/2}^\phi$. The pressure distribution along a cut in the vertical dimension centered at the peak of the pressure in the toroidal dimension is shown in Figure \ref{fig:vertical_cloud_profile}.
It is well described by two Lorentzian distributions separated by 35 mm with equal HWHMs (=$\alpha_{1/2}^Z$). 
The results for the $\alpha_{1/2}^\phi$ evaluations vs. backing pressure are summarized in Figure \ref{fig:half_angle_vs_plenum_pressure}. The measurements were evaluated during the puff ``flattop'', during which time the $\alpha_{1/2}$s are smallest. We plot only $\alpha_{1/2}^\phi$ since they are the ones that matter for constraining the spatial resolution, but note that the $\alpha_{1/2}^Z$ values were very close to the $\alpha_{1/2}^\phi$ values. We note the following:
\begin{itemize}
    \item The degree of collimation improves with increasing backing pressure, as was also observed in \cite{Collis_2006}. The minimum $\alpha_{1/2}$ appears to be $\approx$12$^\circ$.
    \item Also plotted are the half angles measured for the gas distribution from a single unshaped capillary with a 1 mm I.D. It is apparent that our C-D nozzles improve the collimation by roughly a factor of 2.
    \item The half angles in the plane 54 mm from the nozzles are essentially the same as those in the plane at 102 mm. Recall that the center of the FOV is in the plane 110 mm away from the front surface of the nozzle plate.
\end{itemize}

\subsection{\label{sec:heatload}Heat-load management}

We placed the diagnostic's plasma-facing components in locations where heat loading due to plasma conduction and convection was likely to be small. However, radiative loads on those components cannot be avoided, and, as noted in the introduction, all in-vessel components on W7-X must be designed to withstand long radiative emission from the plasma boundary of up to 100 kW/m$^2$. W7-X has infrastructure for extensive water cooling that we used for active cooling of GPI's re-entrant tube. In addition, the nozzle structure was mounted to the water-cooled port liner assembly. Since the nozzle housing is made from the same graphite that is used for the W7-X divertor tiles and the nozzles are SSTL, we depended upon heat conduction to the port liner and radiative loss to keep the peak finite-element modeled temperature of the nozzle structure $\leq700\ ^\circ$C when subjected to a 100 kW/m$^2$ steady-state power flux emanating from the plasma.

Active cooling of the re-entrant tube was provided by 12 mm I.D. tubes that are welded to the outside of the re-entrant tube and carry flowing water. There are six channels distributed around the larger diameter tube that holds the optics, and four channels distributed around the 100 mm I.D. tube on which the turning mirror/shutter is mounted. These two cooling circuits are independently fed by the W7-X cooling water at pressures of $\sim$5 bar and a temperature of $\sim$20 $^\circ$C. Tests and modeling of the six-channel circuit showed that the realized flow rates ($\sim$1.5 m/s) in the different channels are similar, with a full exchange of water every 2-3 s. The cooling tubes run lengthwise and are combined in a collector ring around the vacuum window position on the  larger tube and in a ``cold foot'' on the smaller tube to which the mirror/shutter attaches. The collector ring provides additional local cooling to the window, which is a critical machine safety component. Modeling shows that a 100 kW/m$^2$ steady-state power flux emanating from the plasma and intercepted by the re-entrant tube in the actual geometry increases the water temperature in the circuit by $\sim$5 $^\circ$C for an assumed flow rate of 2 m/s.
The main design difficulty arises from the fact that the component intercepting the most heat flux, i.e., the movable shutter/turning mirror, could not be actively cooled. 

\begin{figure}
    \center
   \includegraphics[scale=0.56]{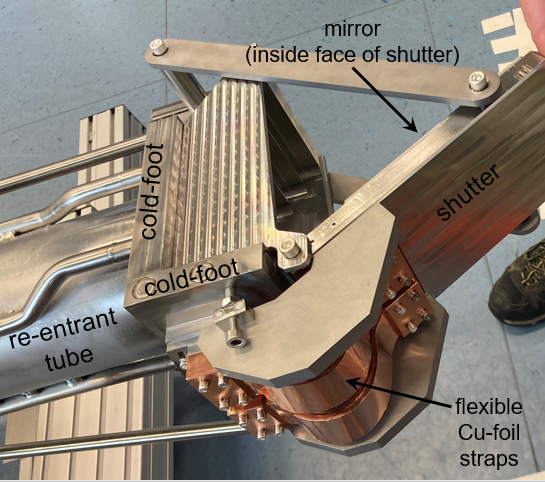}
    \caption{Photo of the front end of the re-entrant tube showing the shutter and the two flexible Cu straps. The cooling water tubes and cold foot are also apparent. The shield covering for the Cu straps (whose inside surface was Cu-plated) has been removed for the photo.}   \label{fig:shutter_with_straps}
 \end{figure}

\begin{figure}
    \center
   \includegraphics[scale=0.15]{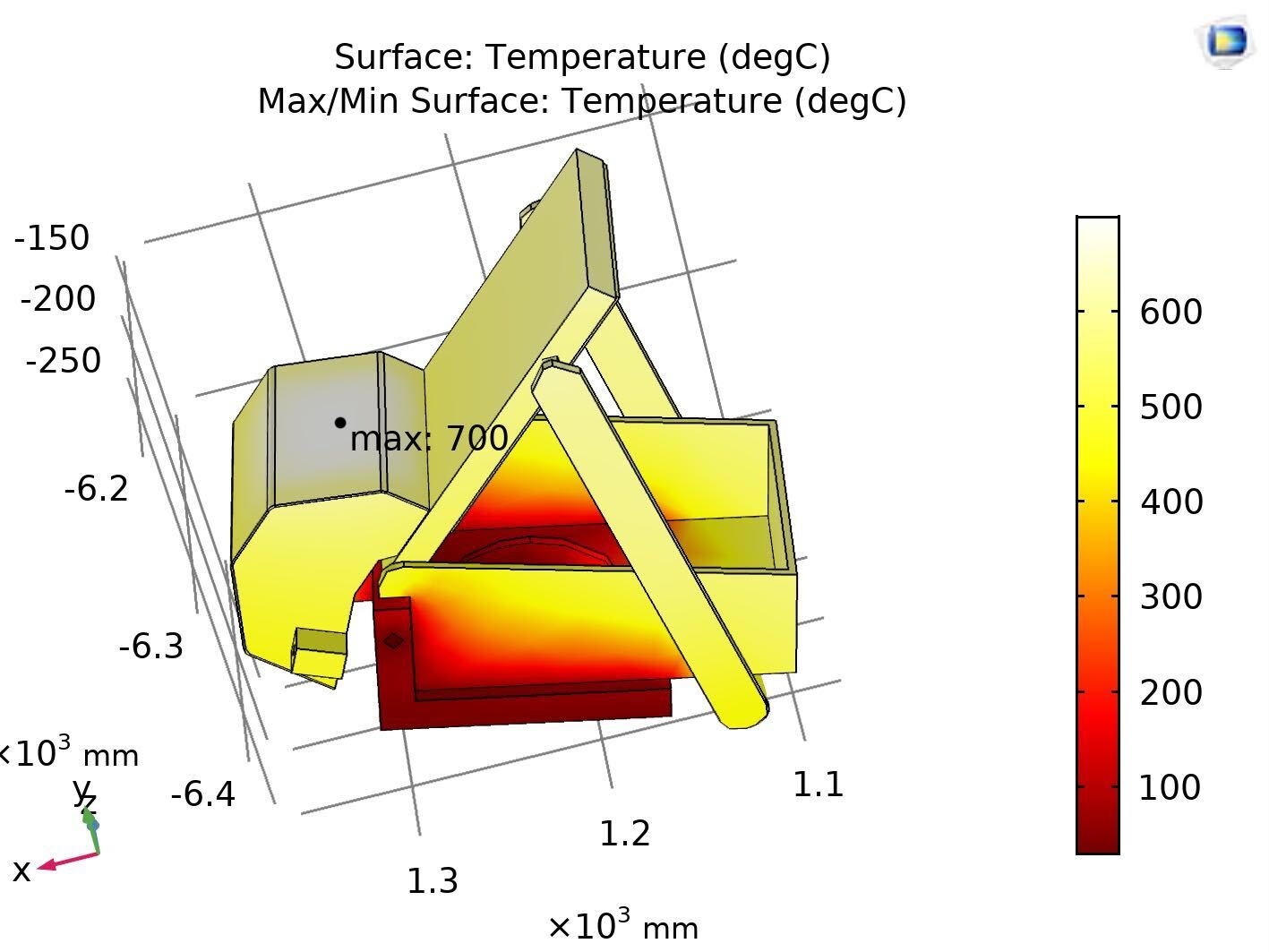}
    \caption{The temperature distribution on the shutter structure when subjected to a 100 kW/m$^2$ radiative heat flux from the plasma boundary as modeled using a finite-element heat transfer code.}   \label{fig:thermal_analysis_of_shutter}
 \end{figure}

The design solution for keeping the shutter/mirror within an acceptable temperature bound was to use flexible ``straps'' that improve the thermal conductivity between the movable shutter and the water-cooled ``cold foot'' described above. The flexible ``straps" are made of stacks of 425 very thin (25 $\mu$m) Cu foils \cite{Thermotive}. A labeled photo of the shutter/mirror component is shown in Figure \ref{fig:shutter_with_straps}. The ``front'' ends of each of the two straps are coupled to a Cu plate sandwiched between the SSTL mirror on the inside of the shutter and the SSTL plasma-facing side of the shutter. Sigraflex pads \cite{Sigraflex} (thin, compressible, high-vacuum-compatible graphite pads with good thermal conduction) are used at each coupling interface. The measured thermal conductivity from the sandwiched Cu plate surface to the cold foot was $\approx$1.0 W/K. With the straps, the finite-element modeled response of the shutter temperature to plasma boundary emission of 100 kW/m$^2$ yields the temperature distribution on the shutter assembly shown in Figure \ref{fig:thermal_analysis_of_shutter}, where the shield covering the straps (not shown in Figure \ref{fig:shutter_with_straps}) reaches a peak steady-state value of $\approx700\ ^\circ$C, and the plasma-facing surface of the shutter reaches $\approx500\ ^\circ$C in the modeling.
The straps reduce the steady-state temperatures by 20-30\% at high power loading but result in a much greater reduction through heat conduction at lower loads where the radiative loss from the components is small.

In order to assess the actual temperatures on the shutter and on the vacuum window, thermocouples were placed on the front face of the shutter and on the quartz window surface just inside the window edge. These temperatures are monitored at all times during W7-X operation.

\section{\label{sec:modeling}Modeling of the expected line emission and detected signal}

\begin{figure}
    \center
   \includegraphics[scale=0.37]{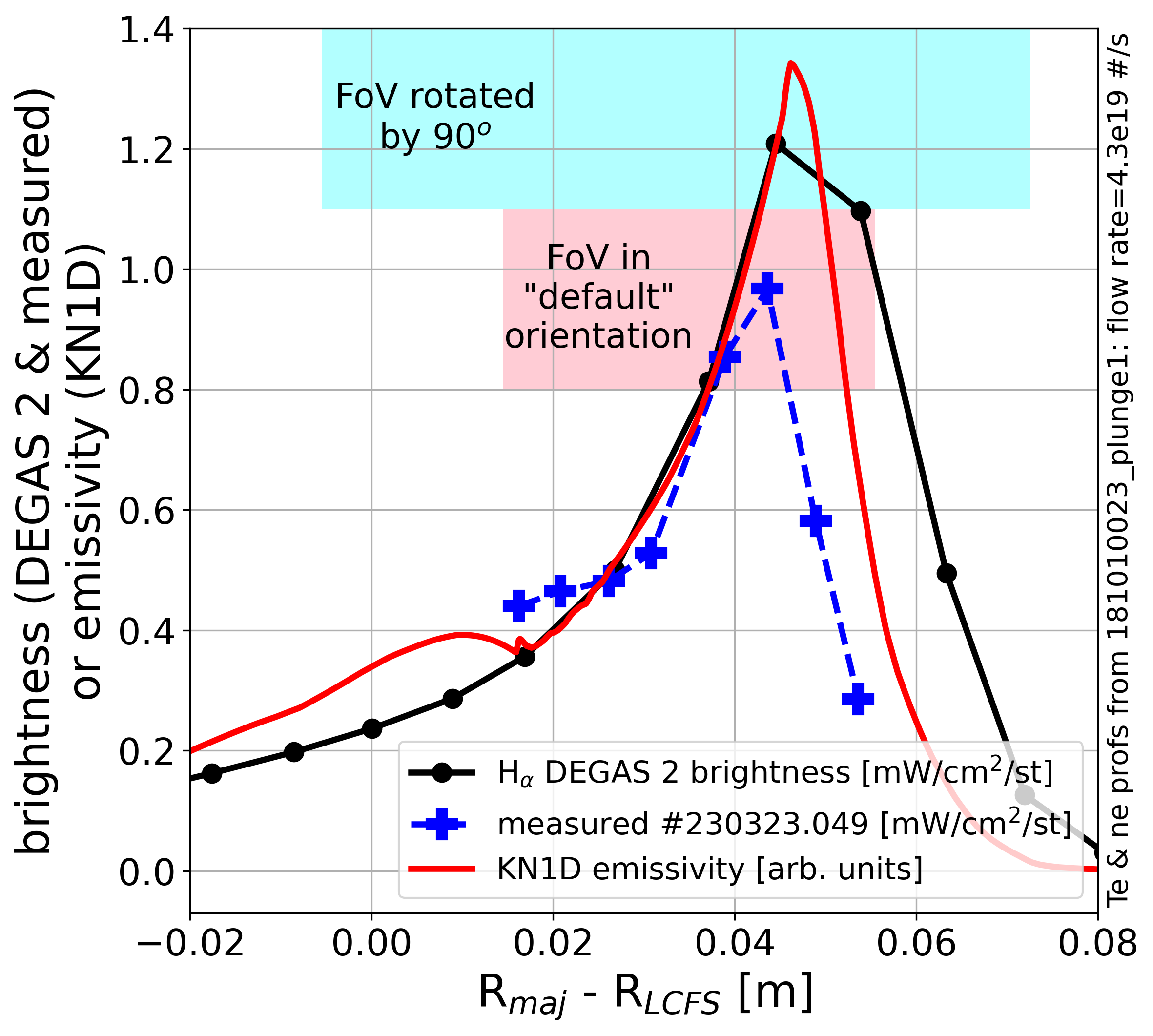}
    \caption{Results of DEGAS 2 and KN1D simulations of H$_2$ flowrates of 4.3$\times10^{19}$ molecules/s into a SOL with ``standard'' configuration n$_e$ and T$_e$ profiles. The black circles show the DEGAS 2-predicted H$_\alpha$ brightness profile in units of mW/cm$^2$/ster. The red curve shows the KN1D H$_\alpha$ emissivity profile scaled to match the DEGAS 2 brightness profile. The blue crosses show the measured time-averaged brightness profile during the first H$_2$ puff into W7-X plasma 20230323.049 scaled by the ratio of the simulation gas flow rate over the actual flow rate for that gas puff (see Section \ref{sec:light signals}). The shaded regions show the extents of the viewed profiles with the FOV with the camera in the ``default'' and ``rotated'' orientations.}   
    \label{fig:DEGAS2_KN1D_Ha_comp}
 \end{figure}

\begin{figure}
    \center
   \includegraphics[scale=0.37]{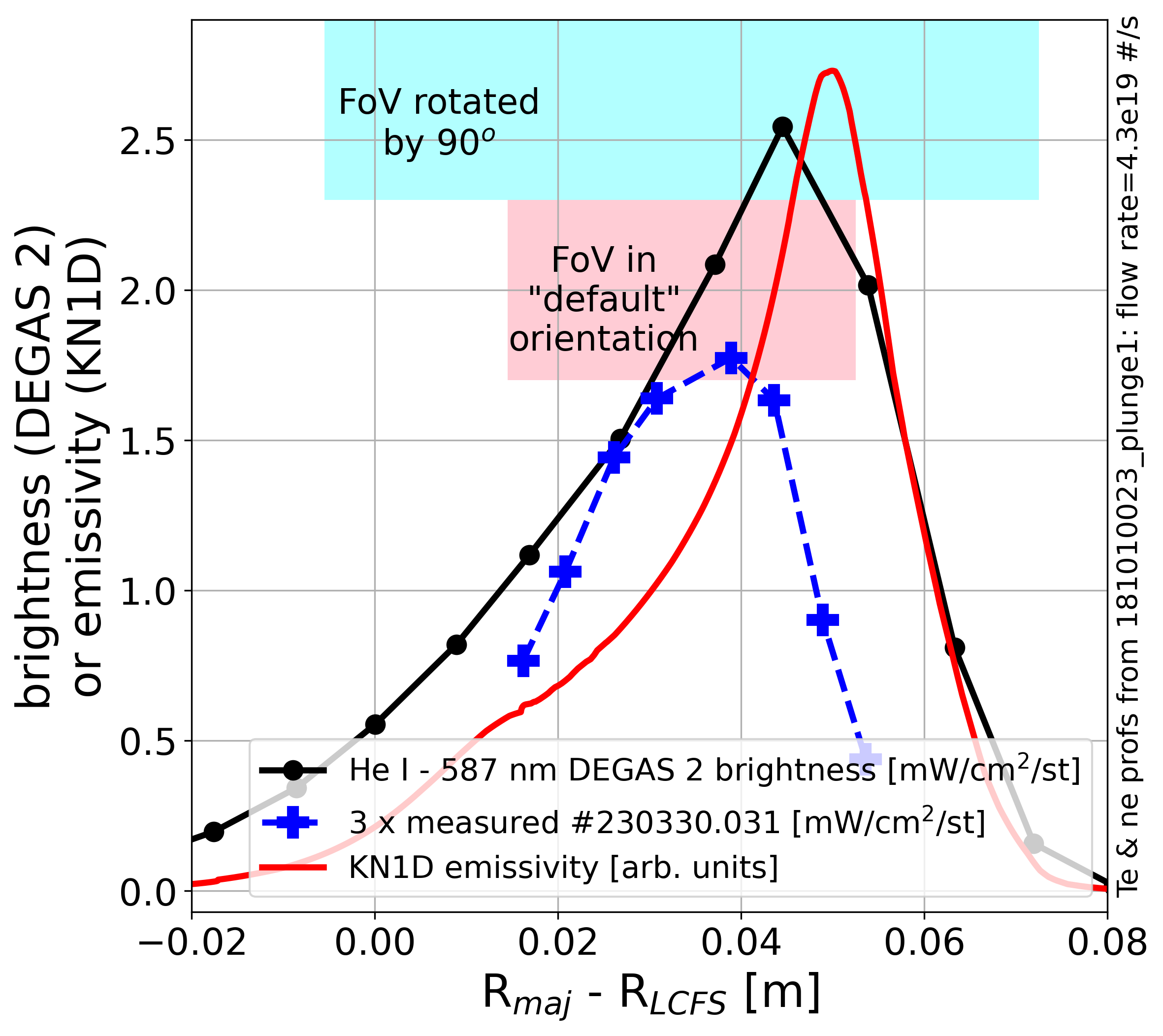}
    \caption{Same as Figure \ref{fig:DEGAS2_KN1D_Ha_comp} but with the HeI - 587 nm line profiles predicted for He flowrates of 4.3$\times10^{19}$ atoms/s.The blue crosses show 3$\times$ the measured time-averaged brightness profile of HeI - 587 nm line at the peak of a He gas puff into  W7-X He-majority plasma 20230330.031 scaled by the ratio of the simulation gas flow rate over the actual flow rate for that gas puff (see Section \ref{sec:light signals}).}   \label{fig:DEGAS2_KN1D_HeI_comp}
 \end{figure}

It was important to model the expected interaction between the W7-X edge plasma and the gas puff to reduce the risk that we would not collect enough light to perform the imaging and to guide the radial placement of the FOV. The modeling tools that were used were the 3D Monte-Carlo neutrals code DEGAS 2 \cite{DEGAS2} and the 1D kinetic neutrals code KN1D \cite{kn1d}. Both codes compute steady-state solutions. DEGAS 2 uses user-input profiles of $T_e$ and $n_e$ with Monte-Carlo neutrals (H$_2$ or He) launched in a user-defined geometry and calculates, among other things, the 3D emissivities of H$_\alpha$ or He I - 587 nm line radiation. A viewing geometry is also input, and chord brightnesses in that geometry are computed from the emissivities. Even though DEGAS 2 is 3D for the neutrals and the emissivities, it uses an axisymmetric target plasma. To adapt this constraint to the non-axisymmetric W7-X geometry, we assumed that the relevant GPI region was small enough it could be approximated by a greatly expanded axisymmetrtic Alcator C-Mod equilibrium, whose equilibrium files were readily available. The LCFS of the expanded equilibrium was placed at a major radius R of 6.015 m and the gas puff was centered vertically at the midplane. Thus, the plasma there was up-down symmetric and not tilted as in the W7-X case (see Figures \ref{fig:FoV} and \ref{fig:average_brightness}). We specified the nozzle spatial locations relative to the nominal LCFS of the equilibrium and the $T_e$ and $n_e$ profiles relative to the same LCFS. We specified the viewing geometry by maintaining the W7-X viewing angles relative to the field lines local to the design FoV at the gas cloud as well as the path lengths for integration of the emission through the gas cloud as they would be on W7-X. The $T_e$ and $n_e$ profiles were approximated using measurements from the W7-X reciprocating probe system in the ``standard'' magnetic configuration\cite{MPM_Nicolai,Killer_NF_2019} made on discharges produced during the previous experimental campaign, OP 1.2. Simulations of H$_2$ and He gas puffs with cloud half-angles of 18$^\circ$ and flow rates of $4.3 \times 10^{19}$ \#/s were performed. The resulting chord brightnesses in the simulated viewing geometry are shown in Figure \ref{fig:DEGAS2_KN1D_Ha_comp} for H$_\alpha$ and Figure \ref{fig:DEGAS2_KN1D_HeI_comp} for the He I (587 nm) line.  

The 1D kinetic neutrals code, KN1D, was run for hydrogen, where it computes the neutral transport, as well as neutrals-plasma interactions (dissociation, ionization, etc.) with stationary 1D radial plasma profiles, and yields, among other things, radial profiles of H, H$_2$, Lyman$_\alpha$ emission, and H$_\alpha$ emission. It considers both the atoms and the molecules. It is run with a boundary condition specifying the radially-inward neutral particle flux at a boundary far from the plasma. It was also run for He, yielding radial profiles of He I atoms and He I 587 nm line emission. We used the same $T_e$ and $n_e$ profiles that were used for the DEGAS 2 simulations. Since KN1D is a 1D simulation and does not consider the cloud shape and viewing geometry, we were primarily interested only in the radial profile of the line emissivities, and those profiles are compared with the brightness profiles from the DEGAS 2 simulations in Figures \ref{fig:DEGAS2_KN1D_Ha_comp} and \ref{fig:DEGAS2_KN1D_HeI_comp}. The KN1D emissivity profiles have been scaled to roughly match the DEGAS 2 brightness profiles. 

We primarily learned two important things from these simulations:
\begin{itemize}
\item The predicted brightnesses over the FOV are in the range $\sim$0.2 to $\sim$2.5 mW/cm$^2$/ster for puff flow rates of $\approx4 \times 10^{19}$ \#/s into  ``standard'' configuration plasmas. Recall (Table \ref{tab:perturbation_flowrates}) that modeling indicated that this flow rate would produce a 10\% density perturbation of a low-density W7-X plasma. 
\item Locating the FOV in its design location places the maximum in the predicted brightness profile near the outside edge of the ``default'' orientation FOV (see Figure \ref{fig:FoV}).
\end{itemize}

Using this brightness range, we estimated the expected signal levels on the detectors in order to ensure that the system is sensitive enough to perform its imaging function. The estimated photon flux onto an on-axis pixel is
\begin{eqnarray*}
    \Gamma_{on-axis}(ph/s)&= B^{ph}R_{m}T_{win}T_{lenses}T_{filt}A_{det}\Omega_{image}\\
&\approx 1.19\times10^{12}\times B^{mW}(\mathrm{for\ H_\alpha})\\
        &\approx 1.06\times10^{12}\times B^{mW}(\mathrm{for\ HeI\ 587\ nm}),
    \label{eq:brightness_estimate}
\end{eqnarray*}
where $B^{ph}$ is the brightness in units of [photons/s/unit area/ster], $B^{mW}$ is the brightness in units of [mW/cm$^2$/ster]; $R_m$ is the reflectivity of the turning mirror; the $T$s are the transmissions of the window, lenses, and filter respectively, and $A_{det}\Omega_{image}$ is the on-axis etendue. The estimate (supplied by the manufacturer) of the APDCAM detector sensitivity at a mid-range bias-voltage of 360 V was: 7$\times10^6 $ph/s corresponds to 1 digitization count. Thus an H$_\alpha$ brightness of 1 mW/cm$^2$/ster with a bias-voltage of 360 V would result in a 170,000 count signal level. Since this is 10.4$\times$ greater than the 14-bit full-scale level, we were assured that the system sensitivity was more than adequate and that a neutral density filter may be required in the system to reduce the light levels. This is addressed in subsection \ref{sec:light signals}.  

\section{\label{sec:performance}Performance of the W7-X GPI System}

The GPI system was installed on W7-X during the 2021-2022 time period between experimental campaigns OP 1.2 and OP 2.1 and commissioned just prior to the start of OP 2.1. Before describing the performance details of the key GPI components and the images obtained, we note that GPI measurements were made on over 80\% of the viable W7-X shots/programs during the OP 2.1 campaign (November 2022 through March 2024), i.e. $\sim$1800 gas puffs distributed over $\sim$800 discharges. 

We address the machine safety issues first. Operationally, we chose to open the shutter at the beginning of a run day, i.e. not cycling it (closed-open-closed) for each program. The maximum temperature increase registered by the shutter thermocouple due to a single plasma program 
was $55\ ^\circ$C in response to a 113 sec long $P_{in}$=5 MW plasma with a 100\% radiation fraction. An initial rapid decay from that temperature increase, followed by a much longer decay time, indicates that the longer decay time is more representative of the bulk temperature of the shutter. Thus, the maximum bulk temperature rise for the shutter is probably significantly lower than 55 $^\circ$C. In any case, heat loads from OP 2.1 plasmas were not a problem for the shutter/mirror structure, implying as well that the radiative fluxes from the OP 2.1 plasmas are much less than 100 kW/m$^2$. This includes the plasma program with 1.3 GJ of input power, a W7-X record to date \cite{grulke_iaea_overview_2023}. Furthermore, the maximum temperature rise of the vacuum window from any W7-X plasma during the campaign, as measured by the window thermocouple, was 1.4 $^\circ$C, and, therefore, also appears to be robust to operational thermal loading. 

Since a key heating system for W7-X is the Electron Cyclotron Resonance Heating (ECH) system (providing up to 7.5 MW) and since the GPI re-entrant tube is in a port adjacent to an ECH launcher port, we were also concerned about a heat load on the fast camera from stray ECH radiation that propagates up the re-entrant tube to the camera. We were not confident in the results from attempts to model this possible heat load on the camera. Thus, during OP 2.1, we measured the ECH heat flux through the aperture immediately in front of the camera using an ECH bolometer provided by the ECH group \cite{Oosterbeek_SOFT_2024}. 
With the GPI shutter open and using the same 113-second $P^{ECH}_{in}$=5 MW plasma discharge that resulted in the 55 $^\circ$C temperature increase on the shutter thermocouple, the stray ECH power flux passing through the aperture just in front of the camera was found to be $\approx 70$\,W/m$^2$, a level much smaller than one that would damage the camera.

In the two following subsections, we review the performance of the GPI system in three crucial areas: the perturbation of the gas puff on the global W7-X plasma, the signal-to-noise in the images, and the spatial resolution of the images.

\subsection{\label{sec:gas-puff}The gas puff}

\begin{figure}
    \center
   \includegraphics[scale=0.45]{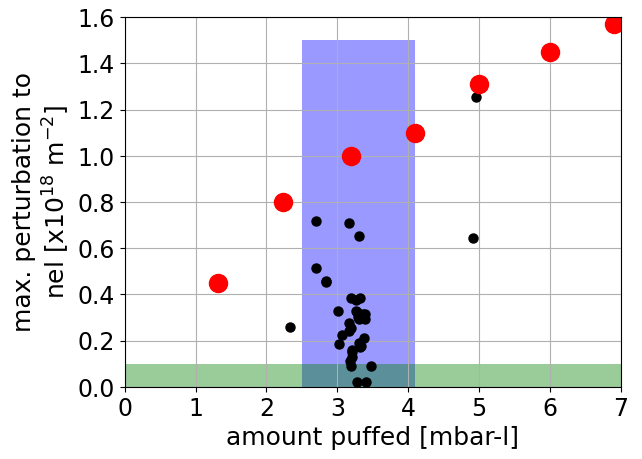}
    \caption{The maximum change in the line-integrated density, $\Delta n_{el}$, measured as a function of the amount of H$_2$ puffed in a GPI single puff. Red circles: perturbation with the density control feedback off; black circles: feedback on. Blue band: typical range of puffed amounts. Light green band: minimum detectable change in $n_{el}$. These evaluations were made on plasmas with $n_{el}$'s that ranged from 2.2 to 10$\times10^{19}$ m$^{-2}$.}   
    \label{fig:max_perturbation}
 \end{figure}

We quantify the perturbation to the W7-X plasma from the GPI gas puff by examining the line-integrated density obtained from the W7-X single channel dispersion interferometer system \cite{brunner2018real}. This measurement does not pass through the localized puff-plasma interaction region. The perturbation analysis is complicated by W7-X's excellent density feedback control that uses the line integrated density, $n_{el}(t)$, as the sensor. The results of the analysis are summarized in Figure \ref{fig:max_perturbation}, where the maximum change in $n_{el}(t)$ due to the puff is plotted vs. the total amount of (H$_2$) gas puffed for both 0.05 s and 0.1 s valve-open durations. The red circles are the results of a plenum pressure scan accomplished during the commissioning phase of the OP 2.1 campaign \textit{before} the density control feedback system was operational. These perturbations in the range of puff amounts that were typically used during the campaign, indicated by the blue shaded region, are $\lessapprox0.12\times 10^{18}\ m^{-2}$. Recalling the perturbation modeling discussed in subsection \ref{sec:nozzle design}, we note that this perturbation is $<$6\% of the lowest density plasmas ($n_{el}\approx2\times 10^{19}\ m^{-2}$) typically run in W7-X and implies a fueling efficiency, $\epsilon_{fuel}$, of roughly 0.4. 

The vast majority of OP 2.1 GPI measurements were done with the density control feedback on. A sub-sample of those for which a perturbation assessment could be made is shown by the black circles in Figure \ref{fig:max_perturbation}. Most of those feedback-on points show a maximum $n_{el}(t)$ perturbation $<0.4\times10^{18}\ m^{-2}$, indicating that the system successfully corrects for the GPI injection by reducing the fueling flow rate, with the correction occurring within roughly 0.3 s. The feedback-on points above that value are from plasmas very early in the campaign when the control system was being tuned up. We do not know the reason for the two feedback-on outliers at 5 mbar\,l. They are from the same run day, and the feedback response looks uncharacteristically slow. In any case, the global perturbations are small, typically $\ll$3\%, and this is a primary reason why the GPI measurements were made on such a large percentage of W7-X shots/programs.

\subsection{\label{sec:light signals}Light signals, images, and spatial resolution }

It was made clear in previous Sections that getting images with excellent signal-to-noise ratio (SNR) was a primary priority. The DEGAS 2 modeling indicated that this would be the case for expected gas puff flow rates, and this proved to be true. 

\begin{figure}
    \center
   \includegraphics[scale=0.57]{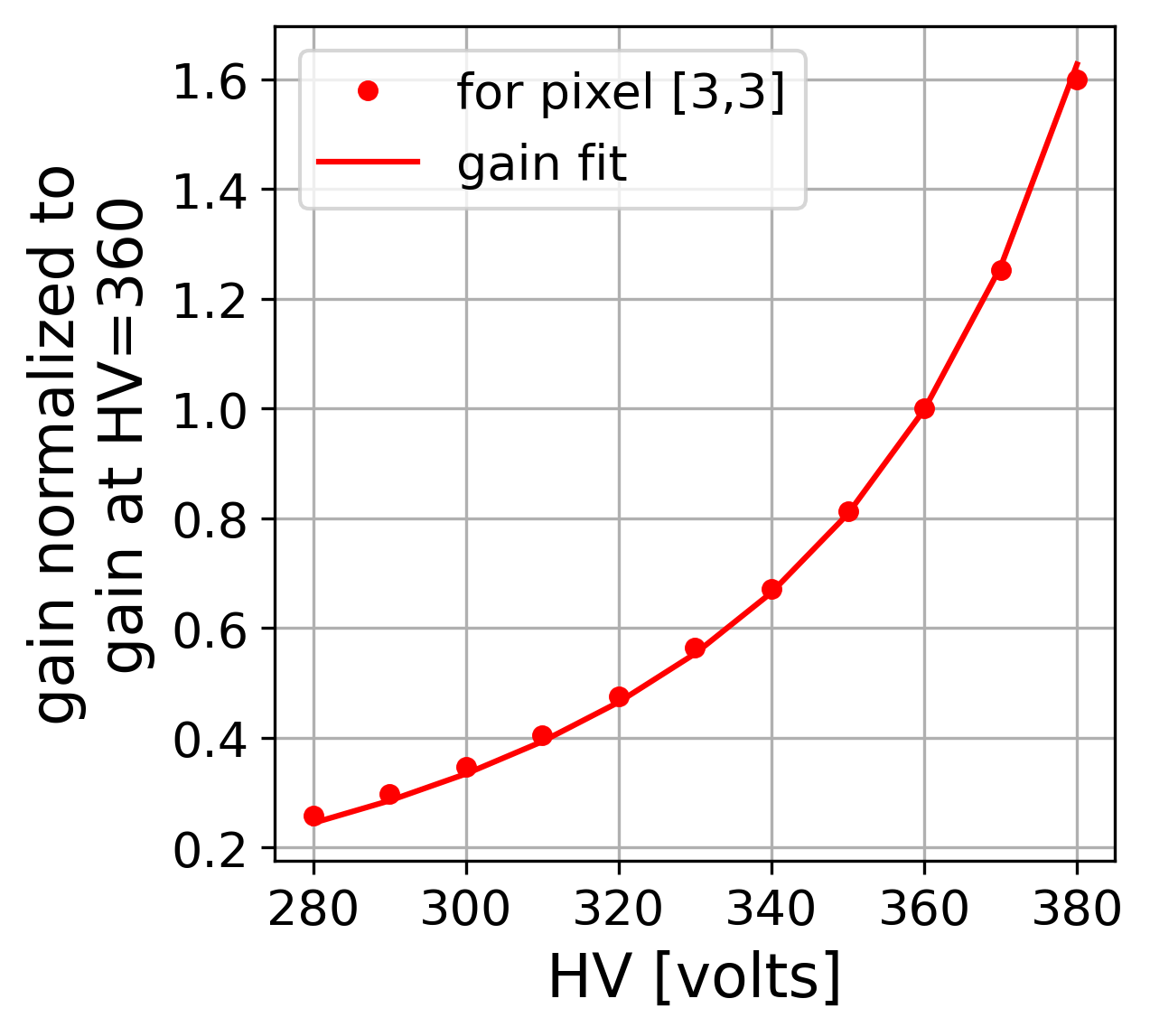}
    \caption{The relative gain measured as a function of applied bias voltage for one of the 8 $\times$ 16 APD detector elements. The gain curves for the other APD elements are very similar. A normalized gain of 1.0 has been chosen for the 360 V bias voltage. The solid line is a fit to the measured values (red dots).}   
    \label{fig:gain_curve}
 \end{figure}

\begin{figure}[hbt!]
    \center
   \includegraphics[scale=0.45]{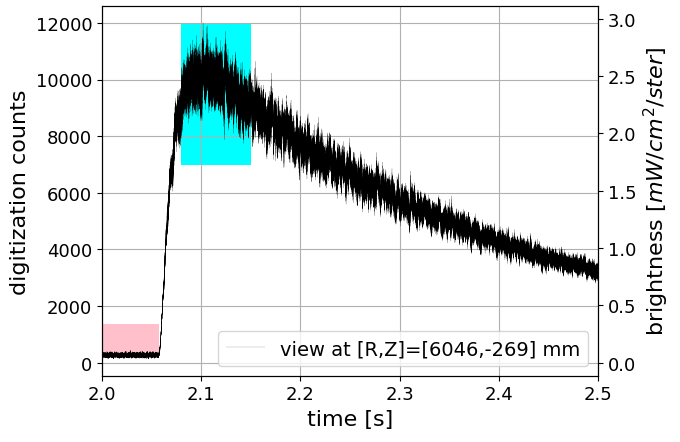}
    \caption{The signal time history from the pixel viewing the labeled [R$_{maj}$,Z] location in the FOV. The absolute H$_{\alpha}$ brightness are given by the y-axis values on the right. The gas from a puff with a 0.05 s valve-open duration and a 1.002 bar backing pressure entered the view at about 2.08 s. The total amount of gas puffed was 3.47 mbar\,l. The 0.07 s interval (cyan) around the maximum brightness is when the ``physics measurements'' are typically made. The interval before the puff (pink) gives the level of intrinsic H$_{\alpha}$ emission.}   
    \label{fig:signal_time_history}
\end{figure}

Almost all of the OP 2.1 results were for H$_2$ puffs into H-majority plasmas, although for two run days, helium was puffed into He-majority plasmas. \textit{Since the puff perturbations were small, we chose to maintain good spatial resolution by keeping the backing pressures/flow rates in mid-range in order to keep the toroidal spread of the gas cloud small} (i.e., with $\alpha_{1/2}\approx14^\circ$ - see Figure \ref{fig:half_angle_vs_plenum_pressure}). Under these conditions, we had to \textit{decrease} the light flux onto the detectors to keep them from saturating. This was typically accomplished using a 5\% neutral density filter in front of the H$_\alpha$ (or He - 587 nm) interference filter. This also allowed us to maintain high signal levels by changing the APD gains in response to changes in light levels due to different plasma conditions. The gain curve from one of the detector pixels is shown in Figure \ref{fig:gain_curve}; a dynamic range of $\approx$6 is evident.
Typically the high voltage for each of the four regions on the APD array was set independently throughout an experiment day so that signals from each region would be as high as possible without incurring any saturation of the 14-bit (16384 count) digitization range. A typical single pixel signal time history is shown in Figure \ref{fig:signal_time_history}. The electronic noise on the digitized signals is 25-30 counts. During the ``physics-measurement" portion of the time history (the cyan-shaded region), the average signal level in the Figure \ref{fig:signal_time_history} example is $\sim$10000 counts (about 60\% of full-scale), yielding an SNR of 370. For plasmas in the ``standard'' configuration, typical SNRs vary from $\sim$50 to $\sim$500 and can vary significantly over a single image. Nonetheless, these typical values are certainly enough to provide very low-noise images. The ``hash'' on the signal during the puff duration is from emission fluctuations due to plasma fluctuations and is exactly what GPI is supposed to measure. 

 \begin{figure}[hbt!]
    \center
   \includegraphics[scale=0.5]{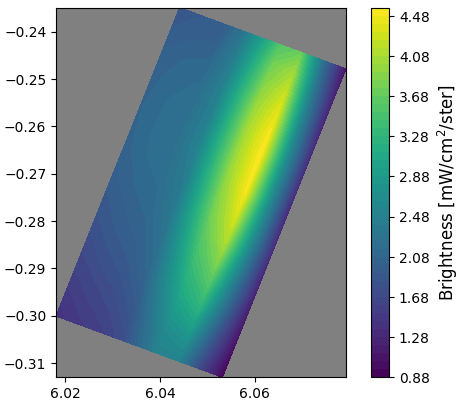}
    \caption{The spatial distribution of the brightness time-averaged over the ``physics-measurement'' interval described in Figure \ref{fig:signal_time_history}. The x-axis/y-axis is the Z (vertical)/R (major radial) coordinate [in m] in the W7-X coordinate system. The GPI system records the images at 2 Msamples/s. A minor radial lineout through the poloidal center of the FOV is the profile used for the comparison made in Figure \ref{fig:DEGAS2_KN1D_Ha_comp}.} 
    \label{fig:average_brightness}
\end{figure}
Brightness levels averaged over the FOV (in the ``standard'' configuration) are typically in the range of 1 to 2.5 mW/cm$^2$/ster, with an example of the spatial variation of the time-averaged brightness shown in Figure \ref{fig:average_brightness}. The absolute brightness calibration comes from a full system calibration on the installed diagnostic using an absolute continuum source placed in the FOV during in-vessel access in 2023 and is not based on the estimates used when evaluating Eq. \ref{eq:brightness_estimate}. The long dimension of the FOV in this camera orientation is essentially poloidal, while the short dimension is radial, i.e. normal to the LCFS. The poloidal distribution of the average brightness is roughly centered in the FOV and varies by about a factor of 2 for a constant radial coordinate. The radial profiles have brightness maxima somewhat inside the outermost edge of the FOV and vary by about a factor of three across the FOV at a given poloidal coordinate. Thus, the emission ``coverage'' throughout the FOV is quite good for the ``standard'' configuration, as was desired in the design. We have investigated the quality of our DEGAS 2 and KN1D modeling predictions by comparing them with actual measured profiles even though the predictions were based on $n_e$ and $T_e$ profiles from 2018 W7-X plasmas. The comparisons (made after scaling the measured brightnesses by the ratio of the simulated to measured flow rates) are shown in Figures \ref{fig:DEGAS2_KN1D_Ha_comp} and \ref{fig:DEGAS2_KN1D_HeI_comp}, with the H$_\alpha$ profile measurement coming from the time-average shown in Figure \ref{fig:average_brightness}. The predictions for H$_2$ puffs are quite good, given that the predictions were made using profiles from a different plasma. The prediction for the He puff is 3$\times$ to 4$\times$ larger than what was measured although the profile shape is reasonable. We do not know the reason for this overestimation; it cannot be blamed on the different $n_e$ and $T_e$ profiles. 

Of course, the radial distributions of the emission change depending on the magnetic configuration, the position of its LCFS relative to the FOV (see Table \ref{tab:FoV_extents}) and the local plasma parameters. There is sufficient emission in all of the common configurations except ``high iota'', where emission from the outer part of the FOV is much reduced and necessitates the removal of the 5\% neutral density filter. 

It is evident in Figure \ref{fig:signal_time_history} that there is a non-zero light signal before the gas puff. That is due to intrinsic H$_\alpha$ emission along the line of sight. It is best for imaging \textit{local} fluctuations that this emission be small compared to the emission due to the gas puff, as it is in the example shown in the Figure. We have examined under which conditions the intrinsic emission may not be sufficiently small. The intrinsic emission is approximately linear with the plasma density but is still small compared to the puff emission as long as the plasma radiative fraction is less than about 60\% and the line-averaged density is less than $\sim 8\times10^{19}$ m$^{-2}$. Above that radiative fraction, the plasmas tend to be detached from the divertor targets, and the puff emission can be small since the radiation front moves radially inwards, out of the FOV. Under those circumstances, the intrinsic emission can be 10 to 50\% of the puff emission and the images will be somewhat compromised. To investigate those plasmas, it is best to puff He gas into H$_2$ majority plasmas, thereby avoiding the intrinsic emission issue completely. However, in the large majority of OP 2.1 W7-X plasmas, this intrinsic emission was very small and did not compromise the images.

We now examine the issue of spatial resolution in the images, estimating the wavenumber range, the minimum distance between two features that can be resolved, and the minimum size of the image of a very thin field-aligned emission filament when viewed in the actual geometry, which we define as the ``instrumental resolution''. Ideally, this latter quantity would be the size of the region viewed by a single pixel, but as will be shown, there is smearing of filament images due to the toroidal extent of the gas cloud.

With the camera in the ``default'' orientation, 16 pixel views are distributed poloidally. The view centers are separated by $\Delta_{pol}\approx4.5$\,mm, resulting in a $k_{pol}$ range of $2\pi/(2L) = 0.45$\,cm$^{-1}\lessapprox k_{pol} \lessapprox6.9$\,cm$^{-1}=2\pi/(2\Delta_{pol})$, where the lower limit is based on the assumption that we can discern one-half of a wavelength using the full poloidal length, $L$, of the FOV. The pixel-view spacing in the radial dimension is non-uniform, but has an average spacing of 5.3\,mm and an estimated $k_{rad}$ range of 0.84\,cm$^{-1} \lessapprox k_{rad} \lessapprox 5.9$\,cm$^{-1}$. To resolve two features of equal brightness as distinct, at least three contiguous pixel views must be examined, with the brightness of the center view being less than that of its neighbors. In our case, this spatial resolution criterion gives a poloidal (radial) resolution of 2$\Delta_{pol}\approx9$\,mm ($2\Delta_{rad}\approx11$\,mm) respectively, as long as the ``instrumental resolution" allows it, as we will examine next.

\begin{figure}
    \center
   \includegraphics[scale=0.35]{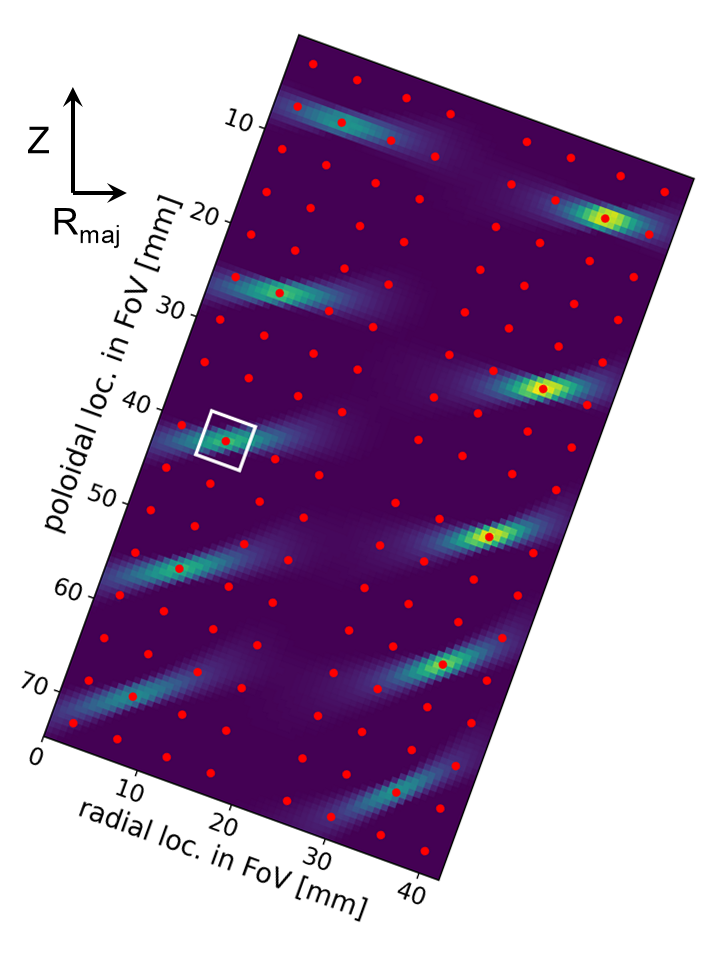}
    \caption{Modeling results of ten 2.4\,mm diameter field-aligned filaments illuminated by the gas cloud mapped onto the ``FOV registration plane'' viewed by the camera in its ``default'' orientation. The x-axis (y-axis) is the distance into the radial (poloidal) dimension of the FOV. The red dots are the centers of the camera pixel views in this plane. The white rectangle is the area viewed by a single pixel.}   
    \label{fig:spatial_resolution}
 \end{figure}

If the gas cloud were a thin sheet in the ``FOV registration plane'', then the sizes, as well as [R$_{maj}$,Z] locations, assigned to features in the images would be ``exact'', limited only by the size of the area viewed by each camera pixel and the accuracy of the spatial calibration. However, the toroidal spread of the gas cloud and the fact that our sight lines are not quite field-aligned within the gas cloud lead to the smearing of field-aligned emission structures in the images. Our analysis of this ``instrumental'' effect relies on our knowledge of the gas puff's double-Lorentzian distribution in planes normal to the nozzle face and knowledge of the sight lines over which the light is collected. Since we are most interested in imaging turbulent structures that are largely field-aligned with $k_\parallel \ll k_\perp$, we model field-aligned filaments. We assume the filament emissivities are proportional to the local gas pressure in the plane parallel to the nozzle face, i.e. the planes in which the distributions were measured. This ignores the reality that those planes are not tangent to the local fluxes surfaces (there is an $\approx 9^\circ$ angle between them), and we presume that this introduces only second-order errors in the analysis. Each modeled filament is assigned a puncture point in the ``FOV registration plane''. We examine brightnesses of a large number of sight lines that image to the detector plane, constructing a high-resolution image of the ``FOV registration plane''. An example using a set of ten simulated filaments, each with a 2.4\,mm diameter circular cross-section, is shown in Figure \ref{fig:spatial_resolution}. It shows the ``instrumental broadening'', with the images of the thin filaments strongly elongated dominantly in the major radial (R$_{maj}$) dimension, leading to significant \textit{radial} smearing, especially in the top half of the FOV. Also shown in the Figure are the centers of the views of the actual detector pixels (red dots) along with the viewed area of a single pixel (the white rectangle). It is clear that the modeled smearing can be larger than the radial size of the area viewed by a single pixel. Combining the extent of this smearing with that of the areas viewed by actual pixels enables us to estimate (below) the ``instrumental resolution'', which must always be greater than or equal to the area viewed by a single pixel. 


We summarize the key findings from our spatial resolution analyses and modeling: 
\begin{itemize}
\item The detector-limited wavenumber resolution is 0.45\,cm$^{-1} < k_{pol} < 6.9$\,cm$^{-1}$ and 0.84\,cm$^{-1} < k_{rad} <  5.9$\,cm$^{-1}$ (with the camera in the ``default" orientation).
\item To resolve two features of equal brightness as distinct in the poloidal (radial) dimension, they must be separated by at least 9 (11) mm (in the ``default" orientation). However, the ``instrumental resolution'' in the top portion of the FOV is such that the 11 mm spatial resolution in the radial dimension is marginal there - see the following. 
\item Modeled images of field-aligned structures show elongation dominantly in the R$_{maj}$ dimension. The FWHM along the long axes of these elongated projections range from $\approx$6 to 10 mm. The presence of such elongated structures in the images may complicate any cross-correlation analysis if they are moving in a direction other than one parallel to either the major or minor axis.
\item In the bottom half of the FOV, the radial and poloidal FWHM of the modeled images range from $\approx2.5$ to 4.7\,mm and $\approx 2$ to 3\,mm, respectively, and thus are essentially the same or smaller than the 4.65 mm dimension of the area viewed by a single pixel in that part of the FOV.
\item Modeled images of field-aligned structures in the top half of the FOV show mostly radial elongation with radial FWHM of $\approx4$ to 9 mm (in the ``default" orientation).
\item Combining the optical resolution with the modeled smearing yields values of ``instrumental resolution'' in the range $\approx 4.6$ to 5.5\,mm in the poloidal dimension and $\approx 6$ to 10 mm in the radial dimension (in the ``default" orientation). The resolutions are better closer to the nozzle, i.e. further out into the SOL.
\item From the modeling, a field-aligned structure piercing the ``FOV registration plane'' at [R$_{maj}$,Z] is imaged with a brightness centroid within the pixel viewing that [R$_{maj}$,Z] point in the ``FOV registration plane''. In other words, the [R$_{maj}$,Z] location assigned to a feature in the images will have pierced that viewed region at the toroidal angle of the ``FOV registration plane''. 
\end{itemize}

\subsection{\label{sec:physics example}Example of 2D dynamics measured by GPI}

\begin{figure}[htp]
    \centering
    \begin{subfigure}{.33\textwidth}
    \centering
    \includegraphics[width=.95\linewidth]{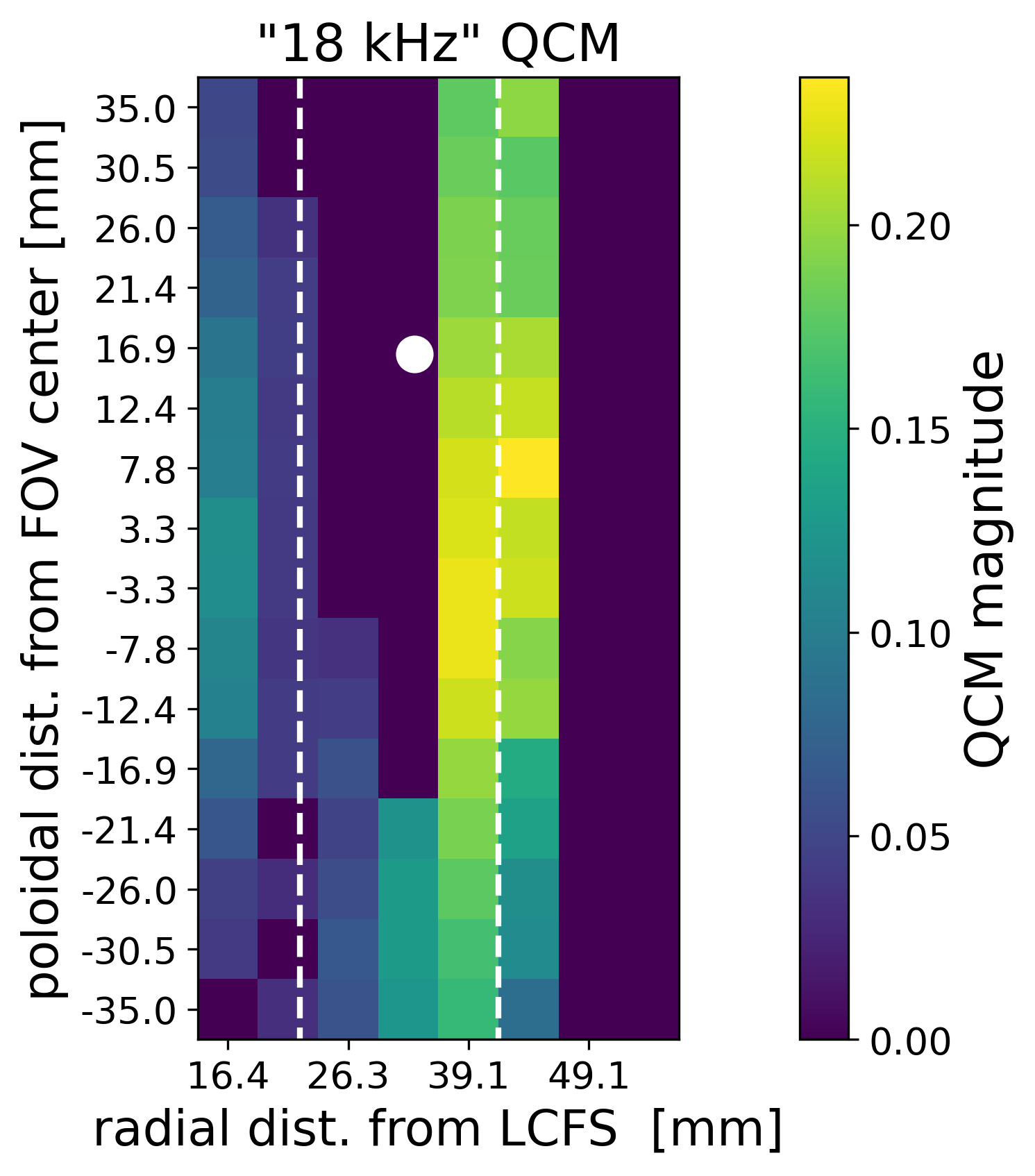}
    \caption{}
    \end{subfigure}%
    \begin{subfigure}{.33\textwidth}
    \centering
    \includegraphics[width=.95\linewidth]{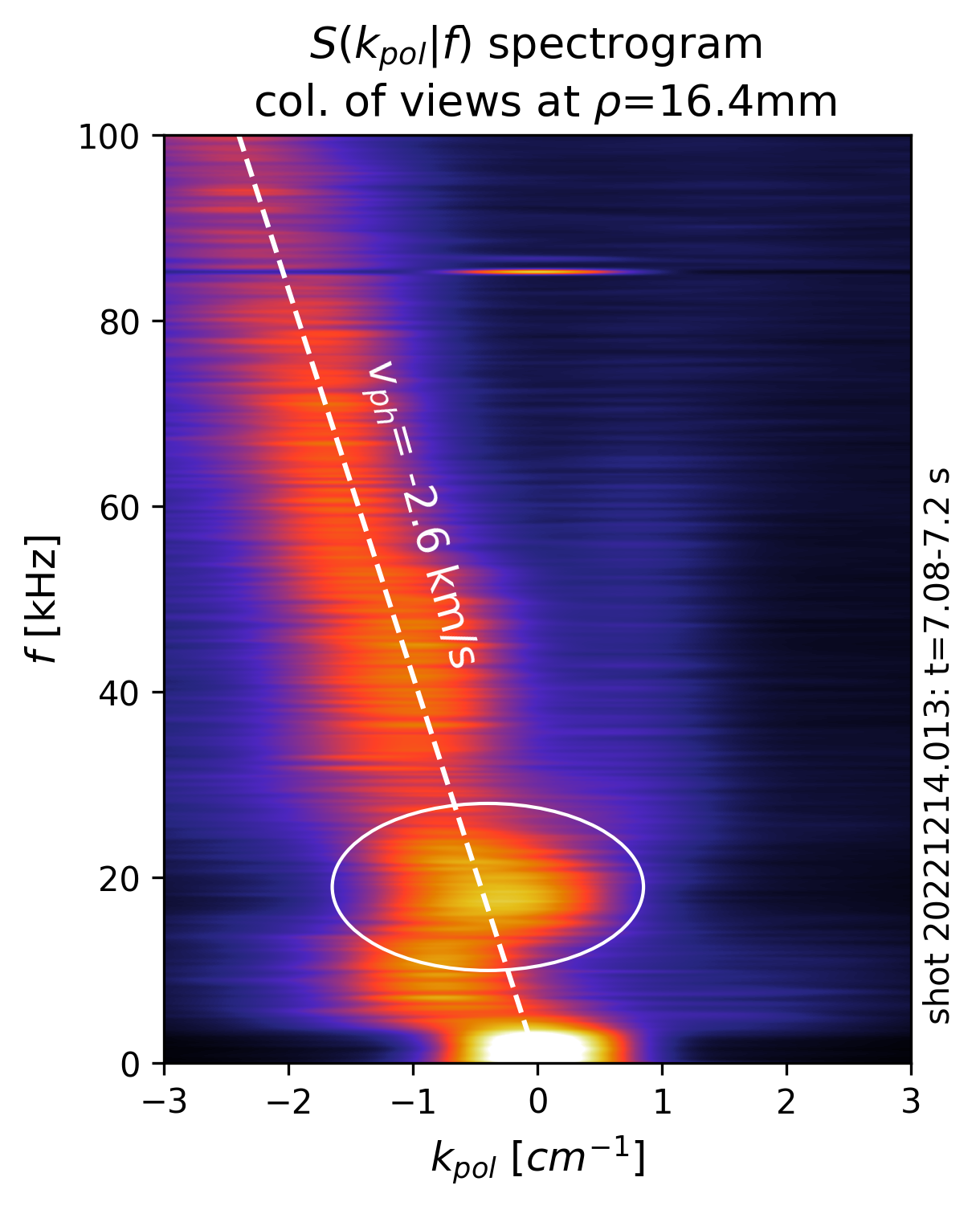}
    \caption{}
    \end{subfigure}%
    \begin{subfigure}{.33\textwidth}
    \centering
    \includegraphics[width=.95\linewidth]{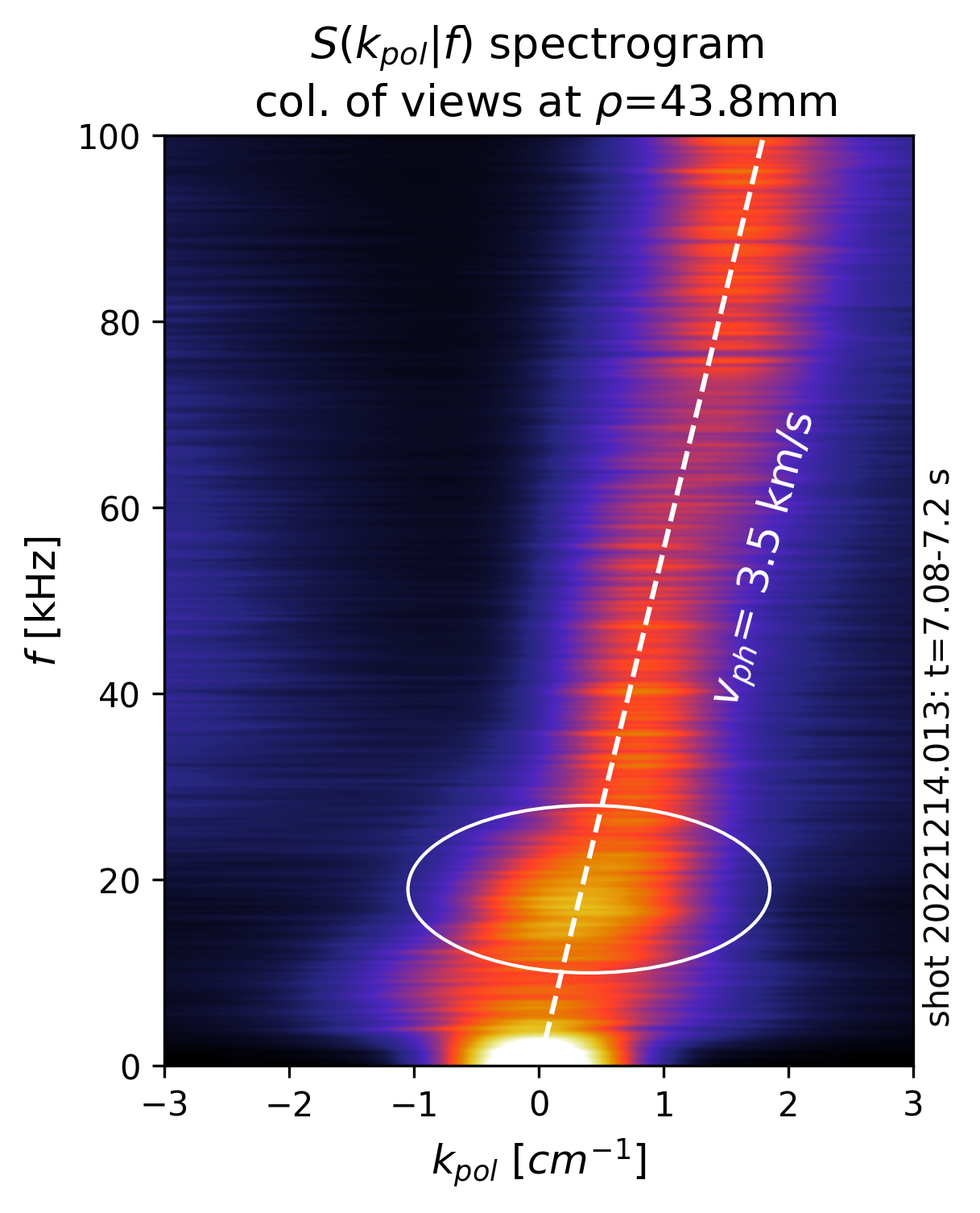}
    \caption{}
    \end{subfigure}%
    \caption{(a) The spatial distribution of the 18 kHz QCM. It is present on the innermost column of views ($\rho$ = 16.4 mm) and on the columns of views with $\rho$ between $\approx36$ and 46\,mm. The color-coded ``QCM magnitude'' is proportional to the integral over frequency of the mode's PSD. (b) A frequency-normalized ($k_{pol}|f$) Fourier spectrogram of the fluctuations in the $\rho=16.4$ mm column of views. (c) Same as in (b) but in the $\rho=43.8$ mm column of views.}
   \label{fig:QCM}
 \end{figure}
We now provide an example illustrating the measurement capabilities of the GPI system for characterizing 2D dynamics of plasma fluctuations in the SOL island region. We investigate a quasi-coherent mode (QCM) present in some W7-X discharges \cite{Han_QCM_W7-X_NF_2019,Zoletnik_W7X_filaments}. In this instance, the OP 2.1 shot/program is 20221214.013. We analyze images from a GPI puff at 7.0\,s when $P_{ECH}=2.0$\,MW, $n_{el}=4.5\times10^{19}$\,m$^{-2}$, and $I_p=2.5$\,kA.  A QCM with a center frequency of $\approx$18 kHz is present in some pixel views but definitely absent in others. The spatial arrangement of the magnitudes of the signals' power spectral densities (PSD) in the mode is shown in Figure \ref{fig:QCM}(a). The mode is present in two radially separated regions. The W7-X magnetic configuration is ``standard'' so the view of the SOL and island is similar to the one shown in Figure \ref{fig:FoV}, but the radial width of the island is significantly larger in this specific case. The island location and size are illustrated in Figure \ref{fig:QCM}(a), where the white solid circle indicates the location of the island O-point, and the white dashed lines show the contour for a 2 km connection length. It is reasonable to conclude that the island plays a key role in the mode location. Examining the dynamics of fluctuations both visually in the videos and analytically, we find the motion is very dominantly in the poloidal direction, which allows accurate evaluations of poloidal wavenumbers and phase velocities using Fourier analysis in the poloidal dimension and time. Frequency-normalized ($k_{pol}|f$) Fourier spectrograms, two of which are shown in Figures \ref{fig:QCM}(b \& c), provide a wealth of information. There is a layer of strong radial shear in the poloidal phase velocities, with the propagation direction changing sign from downward (in the two innermost columns of views) to upward (in the radially outboard columns of views where the QCM reappears). Multiple shear layers within the radial extent of the GPI FOV are not uncommon. Focusing on the QCM, we find it propagating poloidally in the innermost column of views ($\rho=r_{view}-r_{LCFS}$ = 16.4\,mm) at $\approx -2.6$\,km/s (in the lab frame), while in the $\rho=$ 43.8\,mm column of views the phase velocity is $\approx 3.5$\,km/s. The poloidal wavenumbers of the mode are $k_{pol}\approx\pm0.4$ cm$^{-1}$, i.e. approximately GPI's minimum resolvable $k_{pol}$. The mode is moving poloidally along with the broadband fluctuations. The frequency-FWHM of the mode is $\approx7$\,kHz, thus $\Delta_{FWHM}/f_0\approx0.4$. Finally, we note that there is another QCM present in these data - the $\sim$2 kHz mode present in most W7-X plasmas \cite{ballinger2021_1kHz}. It is apparent as the strong very low-frequency feature at $k_{pol}\approx0$ in the Figure \ref{fig:QCM} spectrograms. This brief analysis illustrates some of the phenomena being investigated by GPI on W7-X.

\section{\label{sec:summary}Summary}
In this report, we describe the considerations that went into the design of a new GPI system for the W7-X stellarator. We also describe in detail the different components that make up the W7-X GPI system, and finally, the performance of the realized system. Most important is the result that the system operated reliably during the OP 2.1 experimental campaign, acquiring excellent images for a large fraction of W7-X OP 2.1 plasmas. Detailed analyses of the images will provide an assortment of new and valuable scientific information. We reiterate the key performance results:
\begin{itemize}
\item Excellent SNR in the 2D (radial and poloidal) images acquired at 2 Msamples/s
\item Minimal perturbation to the W7-X plasma by the gas puff
\item Spatial coverage over a $\approx40 \times 75$ mm FOV with a spatial resolution of $\approx$ 5 mm poloidally and $\approx$8 mm radially 
\item A FOV that is typically in the SOL and includes the O-point of an outboard magnetic island in some W7-X configurations and an X-point in another configuration. 
\end{itemize}

It is also important to emphasize some of the unique features of this GPI system:
\begin{itemize}
    \item A ``converging-diverging" nozzle design and fabrication that provides significant collimation of the gas cloud and Mach 1 flow at the nozzle ``throat".
    \item Confirmed collimation of the gas cloud that facilitated placing the nozzle at a safe 110 mm from the center of the FOV. To the best of our knowledge, this separation is the largest of any GPI system to date.
    \item A re-entrant tube of $\sim$2 m length holding a shutter, a turning mirror, a vacuum window, and collection optics that traverses the radial extent of W7-X's superconducting field coils and their cryostat.
    \item A realized design that features robustness to heat loading on plasma-facing components in a long-pulse/steady-state fusion-relevant device. This includes water cooling of the re-entrant tube. Also implemented are a pair of flexible Cu straps that conduct heat from the movable shutter to a water-cooled ``cold foot". We cite the flexible straps as a possible solution for similar future situations on fusion devices.
    \item Direct imaging of the light collected via the turning mirror and lenses onto the fast camera sensor. By building in the capability to rotate the camera about the optical axis of the collection optics, we have registered the locations of two fields-of-view and can switch between them as desired. The ``default" FOV has the long dimension of the view oriented poloidally, while the long dimension of the ``rotated" FOV is radial, i.e. normal to the LCFS.
\end{itemize}

\begin{acknowledgments}
The authors thank Dr. Timo Schr{\"o}der for crucial contributions to the data acquisition process, the entire W7-X staff at IPP Greifswald, and engineer Rui Vieira at the MIT-PSFC. This work was supported in part by the US Department of Energy, Fusion Energy Sciences, Award DE-SC0014251. This work has been carried out within the framework of the EUROfusion Consortium, funded by the European Union via the Euratom Research and Training Programme (Grant Agreement No 101052200 — EUROfusion). Views and opinions expressed are however those of the author(s) only and do not necessarily reflect those of the European Union or the European Commission. Neither the European Union nor the European Commission can be held responsible for them.
\end{acknowledgments}
\begin{parse lines}[\noindent]{#1\\}

The authors have no conflicts to disclose.
\end{parse lines}
\bibliography{aip_style_one_col}

\end{document}